\documentclass[%
 reprint,
 superscriptaddress,
 amsmath,amssymb,
 aps,
 prx
]{revtex4-2}

\usepackage{braket}
\usepackage{graphicx}
\usepackage{dcolumn}
\usepackage{bm}
\usepackage{siunitx}
\usepackage{xcolor} 
\usepackage{tabularx}
\usepackage{dsfont}					
\usepackage{mathrsfs}				
\usepackage{float}
\usepackage{hyperref}

\newcommand{\tr}[0]{\mathbf{tr}}		
\newcommand{\Tr}[0]{\mathbf{Tr}}

\newcommand{\be}[0]{\begin{equation}}	
\newcommand{\ee}[0]{\end{equation}}

\usepackage{hyperref}
\hypersetup{colorlinks, 
	linkcolor={blue!75!black!80!yellow},
	citecolor={blue!75!black!80!yellow}, 
	urlcolor={blue!75!black!80!yellow}
	}

\usepackage[normalem]{ulem}

\begin{document}


\title{Local Fluctuations in Cavity Control of Ferroelectricity}


\author{Jonathan B. Curtis}
\email{jon.curtis.94@gmail.com}
\affiliation{College of Letters and Science, University of California, Los Angeles, CA 90095, USA}
\affiliation{Department of Physics, Harvard University, Cambridge, MA 02138, USA}
\author{Marios H. Michael}
\affiliation{Max Planck Institute for the Structure and Dynamics of Matter, 22761 Hamburg, Germany}
\author{Eugene Demler}
\affiliation{Institute for Theoretical Physics, ETH Zurich, 8093 Zurich, Switzerland}


\date{\today}

\begin{abstract}
    Control of quantum matter through resonant electromagnetic cavities is a promising route towards establishing control over material phases and functionalities.
    Quantum paraelectric insulators\textemdash materials which are nearly ferroelectric\textemdash are particularly promising candidate systems for this purpose since they have strongly fluctuating collective modes which directly couple to the electric field.
    In this work we explore this possibility in a system comprised of a quantum paraelectric sandwiched between two high-quality metal mirrors, realizing a Fabry-Perot type cavity.
    By developing a full multimode, continuum description we are able to study the effect of the cavity in a spatially resolved way for a variety of system sizes and temperatures.
    Surprisingly, we find that once a continuum of transverse modes are included the cavity ends up suppressing ferroelectric correlations. 
    This effect arises from the screening out of transverse photons at the cavity boundaries and as a result is confined to the surface of the paraelectric sample.
    We also explore the temperature dependence of this effect and find it vanishes at high temperatures, indicating it is a purely quantum mechanical effect.
    We connect our result to calculations of Casimir and Van der Waals forces, which we argue are closely related to the dipolar fluctuations in the quantum paraelectric.
    Our results are based on a general formalism and are expected to be widely applicable, paving the way towards studies of the quantum electrodynamics of heterostructures featuring multiple materials and phases. 
\end{abstract}

\pacs{}

\maketitle

\section{Introduction}

Optically engineering properties of quantum materials may potentially allow for the design and development of novel technologies and the creation of phases of matter which are otherwise difficult to obtain and study. 
Ideally, one would like to think of this as adding a new ``control knob" to the toolbox of solid-state physics just like temperature, pressure, external field, and twist-angle, allowing for new explorations of physical systems and device structures~\cite{Basov.2017}. 
For instance, intense electromagnetic radiation can induce nonequilibrium phases of matter and generate new phase diagrams, with sometimes no counterpart in equilibrium~\cite{Oka.2009,Kitagawa.2010,Lindner.2011,Cavalleri.2018,Gao.2020,MariosM22,MariosM20,Pavel20,PashaBiplasmons,Baumann.2010,Babadi.2017,Chaudhary.2020}.
However, the nonequilibrium route towards optical control has a number of drawbacks which impede its practical application, chief among them are problems related to heating, optical access, and complicated theoretical modeling.
Therefore, it would be desirable to obtain a similar degree of optical control without leaving thermal equilibrium.
Recently, it has been argued that this may be done by instead shaping the environment of electromagnetic fluctuations through the use of optical cavities, resonators, and metamaterials~\cite{Bloch.2022,Schlawin.2022}.
Many systems have recently been proposed to be amenable to control in this way, including superconductors~\cite{Sentef.2018,Curtis.2019,Schlawin.2019,Gao.2020}, excitonic insulators~\cite{Mazza.2019}, antiferromagnets~\cite{Curtis.2022,Parvini.2021}, spin-liquids~\cite{Chiochetta.2021}, semiconductors~\cite{Amelio.2021}, quantum Hall fluids~\cite{Paravicini.2019}, and ferroelectrics~\cite{Ashida.2020,Latini.2021,Zhang.2019,Lenk.2022,Lenk.2022b}, with a great deal more proposed to exhibit strong coupling between material and optical excitations~\cite{Basov.2020,Juraschek.2021}.
Recent experiments on the metal-insulator transition in 1$T$-TaS$_2$ even seem to have seen promising signatures of cavity control on the transition temperature~\cite{Jarc.2022}.
Experiments have also seen fascinating phenomena occur when unconventional superconductors are strongly coupled to the quantum electromagnetic bath~\cite{Thomas.2019,Thomas.2021}. 

Ferroelectrics are particularly promising candidates since the relevant fluctuations\textemdash phonon polaritons\textemdash directly couple strongly to the electromagnetic field via the electric dipole transition even down to atomic scale~\cite{Rivera.2019,Berte.2018,Dai.2019,Juraschek.2021}.
Furthermore, there are a number of appealing candidate systems, such as SrTiO$_3$~\cite{Yamanaka.2000,Rowley.2014,Kozina.2019,Palova.2009,Esswein.2022} and various moir{\'e} and Van der Waals materials~\cite{Zheng.2020,Stern.2021,Yasuda.2021,Woods.2021,Moore.2021} which may be suitable for proof-of-principle experiments.
Intrinsic SrTiO$_3$ is believed a quantum paraelectric (QPE)~\cite{Yamanaka.2000,Rowley.2014,Kozina.2019,Palova.2009,Esswein.2022}, lying right at the border of the ferroelectric phase, with long-range order suppressed by quantum fluctuations.
Strain, chemical, and isotope substitution have all been shown to tip the system over the edge in to the ordered phase~\cite{Rowley.2014}, and recently resonant optical excitation of the lattice~\cite{Katayama.2012,Kozina.2019} have also been shown to seemingly induce a transition into the ordered phase~\cite{Nova.2019,Li.2019ilv}, making this a prime candidate for demonstrating cavity control over the phase diagram~\cite{Ashida.2020,Latini.2021,Lenk.2022}.
Previous theoretical investigations have largely been limited to single-mode, dipole coupling, and translationally invariant approximations.
In this Article we show that going beyond these simplifying assumptions can lead to a qualitatively different behavior making our approach necessary in order to describe realistic experiments~\cite{Ashida.2021b}.

In this work, we extend our analysis of fluctuating quantum paraelectric to a fully multimode~\cite{Amelio.2021}, spatially resolved system beyond the standard dipole approximation \textemdash the standard optical approximation which treats the electromagnetic field as homogeneous across the sample.
In fact, this is a crucial technical development, giving a number of predictions which starkly differ from previous simplified models.
By making use of connections to the study of Casimir-Polder and Van der Waals forces, we are able to efficiently solve the full multimode problem including a continuum of electromagnetic modes.
In doing so, we find that in fact the presence of the cavity suppresses ferroelectric fluctuations in the system\textemdash the opposite of what is expected based on a single- or few-mode calculations.
This surprising result then has important implications for future experiments on cavity control of ferroelectricity and potentially other phases of matter~\cite{Jarc.2022,Thomas.2019,Thomas.2021}.

The key insight is using the fluctuation dissipation relation to reformulate the problem in terms of the dielectric response and its variational dependence on material parameters, thereby encapsulating the effect of electromagnetic fluctuations in terms of well-known frequency-dependent electric-field correlation function. 
The behavior of this correlations function is very well studied, dating back to seminal work on the Casimir force~\cite{Casimir.1948,Lifschitz.1956,Dzyaloshinskii.1961,Kenneth.2006} and is by now well documented and experimentally verified.
In fact, cavity control over the QPE fluctuations in a material is closely related to the problem of using the cavity to modify the Van-der Waals forces between virtual dipolar excitations in the cavity~\cite{Philbin.2022}.
Furthermore, we argue our technique can be easily extended to include phonon loss, anisotropy, and mode couplings provided the dielectric constant dispersion $\epsilon(\omega)$ is known well enough, and may even be extended to include more complicated heterostructure geometries such as interfaces between quantum paraelectrics and metals, air, or more exotic two-dimensional systems via characterization in terms of the reflection coefficients at interfaces~\cite{Kenneth.2006}. 
In a rough sense, the problem is similar to calculating the Casimir force but instead of computing the energy as a function of the boundary separation, we keep the boundary conditions fixed and directly compute the photon and phonon fluctuations inside the cavity. 

\begin{figure}
    \centering
    \includegraphics[width=\linewidth]{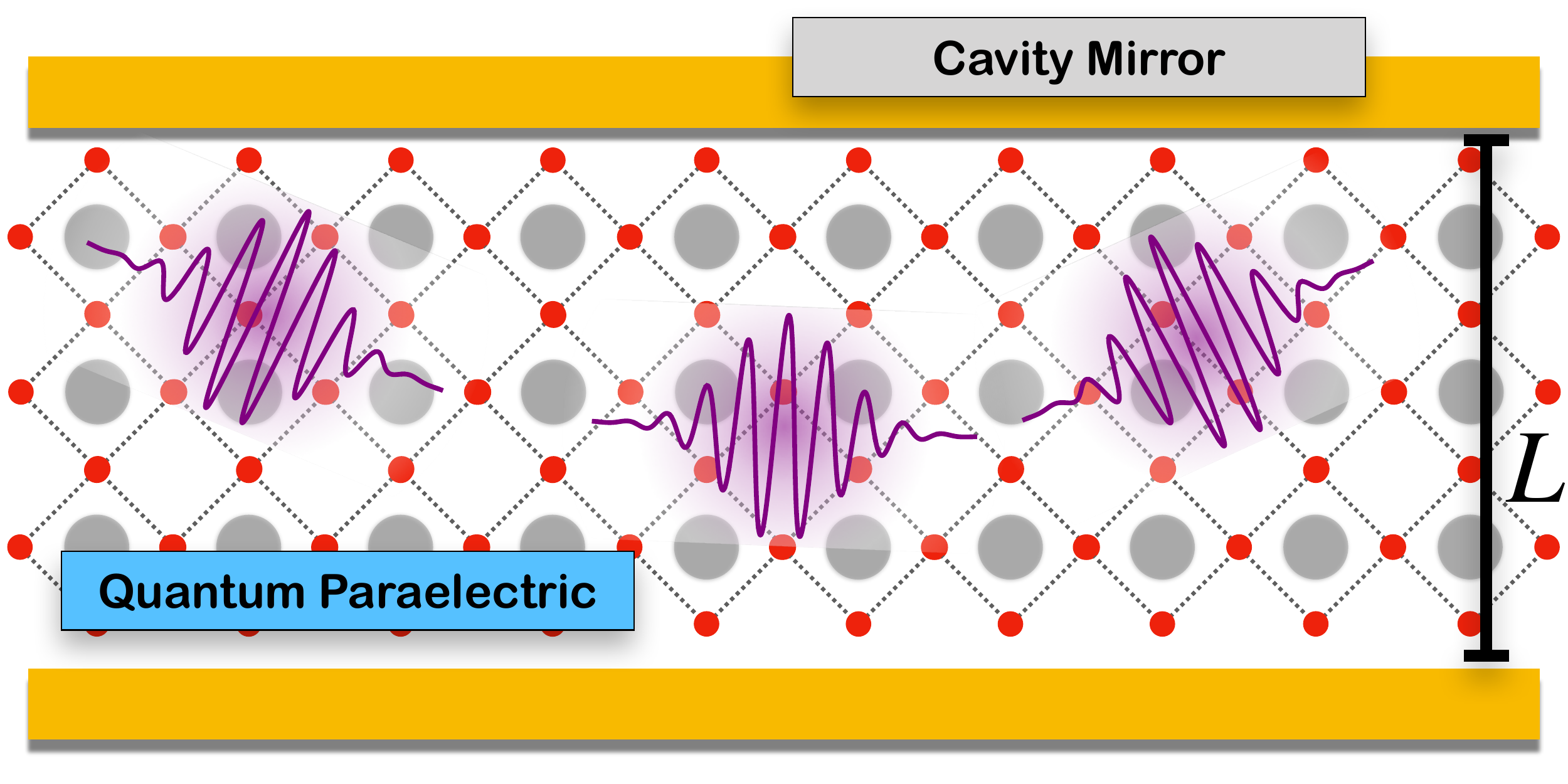}
    \caption{Schematic depiction of cavity. 
    Two perfect metal plates are located in the $xy$ plane at $z=\pm L/2$, and the electromagnetic field and phonon modes coexist within the interior.
    The system has translational symmetry in the $xy$ plane, which leads to a conserved in-plane momentum $\mathbf{q}$, which can be taken along $\mathbf{q} \parallel\hat{\mathbf{e}}_x$, and obeys ideal metallic boundary conditions at $z=\pm L/2$.
    Due to metallic boundary conditions, the photonic part of the wavefunction is suppressed near the boundary, leading to enhanced phonon fluctuations.}
    \label{fig:schematic}
\end{figure}

After obtaining these general relations, we use our method to compute the local behavior of the QPE fluctuations by considering a Fabry-Perot type system with perfect metallic boundaries sandwiching a QPE, illustrated in Fig.~\ref{fig:schematic} in a cross-sectional view.
We find as our key result that actually towards the boundaries of the sample the phonon fluctuations $\langle \mathbf{Q}^2(\mathbf{r},t)\rangle$ increase, resulting in a blue-shift of the soft-mode transverse frequency $\Omega_T$ due to the anharmonic coupling of the phonon mode, characterized by the local displacement field $\mathbf{Q}(\mathbf{r},t)$. 
This leads to an overall thickness dependence which may be pronounced for thinner cavities and results in a diminished low-frequency dielectric constant $\epsilon(0)$, signaling a hardening of the soft polar mode. 

We also find that this effect\textemdash the difference between the surface and bulk fluctuations\textemdash vanishes at higher temperatures, indicating the origin of this effect is of a truly quantum origin, and should drop off once $T \gtrsim \hbar \Omega_T/K_B$. 
For materials such as SrTiO$_3$, with $\Omega_T \sim 2$THz, this provides an important ceiling on the temperature of experiments which is roughly of order 100K. 

The remainder of our Paper is structured as follows.
In Sec.~\ref{sec:fdt} we use the fluctuation-dissipation theorem to connect the phonon fluctuations to the dielectric response of the material.
We first do this in real-time formalism in Sec.~\ref{sub:real-time}, followed by a reformulation on the Matsubara axis for finite-temperature calculations in Sec.~\ref{sub:matsubara}.
In Sec.~\ref{sec:homogeneous} we demonstrate how our results connect to the more familiar method based on phonon-polaritons in the case of a bulk translationally invariant system.
In particular, in Sec.~\ref{sub:high-temp} we show how at high-temperatures the photons decouple highlighting that the cavity control is a manifestation of quantum effects.
In Sec.~\ref{sec:fabry-perot} we apply this technique to derive the local QPE fluctuations in a Fabry-Perot geometry which does not preserve translational symmetry. 
We then conclude by discussing general aspects of our results beyond our cavity-QPE model and highlighting potential directions for future study in Sec.~\ref{sec:conclusion}.

\section{Fluctuation-Dissipation Theorem}
\label{sec:fdt}

In the following we will focus on the case of a local~\footnote{By this, we mean that the phonon dispersion can be safely approximated as flat over the region of interest in momentum space, and therefore the eigenmodes can be localized in real space.}, isotropic, polar phonon mode.
In particular, since the phonon group velocity is much slower than the photon, the approximation of a non-dispersion phonon mode should be suitable for studying electrodynamic effects. 
Going beyond this approximation to include the phonon dispersion would be an interesting direction for future studies and may be important very close to the ferroelectric critical point or in the ordered phase, which we will not study in this work.

We thus consider a model of a local polar phonon mode $\mathbf{Q}(\mathbf{r})$ with transverse optical (TO) mode frequency $\Omega_T$ and effective charge $\eta$ coupled to the electromagnetic field, described by $\mathbf{E}$ and $\mathbf{B}$. 
This system is most simply described in terms of a Lagrangian which generates the equations of motion.
We have 
\begin{equation}
    \mathcal{L} = \mathcal{L}_{\rm EM} + \mathcal{L}_{\rm ph} + \mathcal{L}_{\rm int}.
\end{equation}
The Maxwell Lagrangian is 
\begin{equation}
    \mathcal{L}_{\rm EM} = \frac12 \mathbf{E}^2 - \frac12 \mathbf{B}^2.
\end{equation}
In terms of the gauge potentials, the electromagnetic fields are expressed as
\begin{subequations}
\begin{align}
& \mathbf{E} = -\nabla A_0 - \partial_t \mathbf{A} \\
& \mathbf{B} = \nabla \times \mathbf{A} .
\end{align}
\end{subequations}
Here and throughout we use units where $\hbar = c = k_B = \epsilon_0 = 1$.

The phonon Lagrangian, in the absence of dispersion, is purely local and is simply given by: 
\begin{equation}
    \mathcal{L}_{\rm ph} = \frac12\left[ \left(\partial_t \mathbf{Q}\right)^2 - \Omega_0^2\mathbf{Q}^2 \right] - \frac{\lambda}{4}\left(\mathbf{Q}^2 \right)^2 ,
\end{equation}
where $\Omega_0$ is the bare TO mode frequency, and $\lambda$ is the phonon-phonon interaction strength.
Due to symmetry, these are the only local terms allowed at up to quartic order and second order in time-derivatives.

Finally, we have the dipole-coupling between the phonons and the electric field. 
The phonon displacement field $\bf Q$ generates a polarization which then couples to $\bf E$, via 
\begin{equation}
    \mathcal{L}_{\rm int} = +\eta \mathbf{Q}\cdot\left[-\nabla A_0 - \partial_t \mathbf{A} \right].
\end{equation}
Here, $\eta$ is the light-matter interaction constant and it sets, among other things, the size of the splitting between the longitudinal optical (LO) and transverse optical frequency splitting due to the Coulomb part of the electromagnetic interaction (the so-called LO-TO splitting). 
Before proceeding, in order to perform calculations we must fix a gauge. 
In this work, we will employ the "Weyl gauge", which is obtained by demanding $A_0 = 0$.

We proceed by treating the phonon nonlinearity in the Hartree approximation, such that the system is essentially linear, albeit with a renormalized phonon frequency of 
\begin{equation}
    \Omega_T^2 = \Omega_0^2  + \lambda \langle \mathbf{Q}^2 (\mathbf{r},t)\rangle.
\end{equation}
In general, we allow for spatially varying fluctuations of $\mathbf{Q}$, and therefore the phonon frequency may be renormalized in an inhomogeneous way, which is the subject of this investigation. 

Therefore, our primary objective is to compute the spatially resolved phonon fluctuations, $\langle \mathbf{Q}^2(\mathbf{r},t) \rangle$.
We do this by the familiar linear-response formalism, obtaining the equilibrium fluctuations of $\mathbf{Q}$ by solving for the causal response to an external perturbation. 
We thus introduce a source field $\mathbf{F}(x)$ which couples to the phonon mode via 
\begin{equation}
    \mathcal{L}_{\rm source} = \mathbf{F}(\mathbf{r},t)\cdot\mathbf{Q}(\mathbf{r},t),
\end{equation}
such that 
\begin{equation}
\mathbb{\hat{D}}^R(x,x')=-\frac{\delta \langle \mathbf{Q}(x) \rangle }{\delta\mathbf{F}(x')}\bigg|_{\mathbf{F} = 0} = -i\theta(t-t')\langle [ \mathbf{Q}(x) ,\mathbf{Q}(x')] \rangle
\end{equation}
for causal response function.
Here and throughout, when confusion is not likely, we will use $x = (\mathbf{r},t)$ to represent a spacetime four-coordinate, while $\mathbf{r}$ as a spatial three-coordinate.
From this we can apply the fluctuation-dissipation relation~\cite{Rammer.2007} to obtain 
\begin{equation}
    \langle \mathbf{Q}(x)\cdot\mathbf{Q}(x)\rangle = -\int \frac{d\omega}{2\pi} \coth \frac{\beta\omega}{2}\Im \left[\tr \mathbb{\hat{D}}^R(\mathbf{r},\mathbf{r};\omega)\right].
\label{eqn:FDT}
\end{equation}
This then allows to characterize the {\bf local} density of phonon fluctuations.

\subsection{Real-Time Equations of Motion}
\label{sub:real-time}
The relevant equations of motion can be written down, including the source term which acts on the phonon field. 
This gives us the equations in the frequency domain 
\begin{subequations}
\begin{align}
    & \left[ -\omega^2 + \Omega_T^2 \right] \mathbf{Q}(\mathbf{r},\omega) = \eta \mathbf{E}(\mathbf{r},\omega) + \mathbf{F}(\mathbf{r},\omega) \\
    & +i\omega \mathbf{B}(\mathbf{r},\omega) = \nabla\times\mathbf{E}(\mathbf{r},\omega) \\
    & -i\omega \left[ \mathbf{E}(\mathbf{r},\omega) + \eta \mathbf{Q}(\mathbf{r},\omega) \right] = \nabla\times\mathbf{B}(\mathbf{r},\omega) 
\end{align}
\end{subequations}
We compute the response function of $\mathbf{Q}$ as follows. 
Let us first introduce the bare response function for $\mathbf{Q}(\mathbf{r},\omega)$ of 
\begin{equation}
    \chi_0(\omega) = \frac{1}{-\omega^2 + \Omega_T^2},
\end{equation}
such that we can obtain the response of $\mathbf{Q}$ as 
\begin{equation}
    \mathbf{Q}(\mathbf{r},\omega) = \chi_0(\omega) \left[\eta \mathbf{E}(\mathbf{r},\omega) + \mathbf{F}(\mathbf{r},\omega) \right].
\end{equation}
Our job is not done because we need the response of the phonon not to the {\bf total} force, which is $\eta \mathbf{E} + \mathbf{F}$, but only to the {\bf external} force $\mathbf{F}$, which is partly screened by the electromagnetic field. 

This screening is found by solving the equations of motion for the electromagnetic field. 
We have 
\begin{subequations}
\begin{align}
    & +i\omega \mathbf{B}(\mathbf{r},\omega) = \nabla\times\mathbf{E}(\mathbf{r},\omega) \\
    & -i\omega \left( \mathbf{E}(\mathbf{r},\omega) + \eta \chi_0(\omega)\left[\eta \mathbf{E}(\mathbf{r},\omega) + \mathbf{F}(\mathbf{r},\omega) \right] \right) = \nabla\times\mathbf{B}(\mathbf{r},\omega) 
\end{align}
\end{subequations}
The second equation contains the dependence on the forcing field; we can eliminate the magnetic field to derive a closed equation for the response of $\mathbf{E}$, which we use to find the depolarizing field, from which we find the effective force acting on the phonon mode due to the external force. 

We get 
\begin{equation}
    \omega^2 \epsilon(\omega)\mathbf{E}(\mathbf{r},\omega) - \nabla \times \nabla \times \mathbf{E}(\mathbf{r},\omega) = -\omega^2 \eta \chi_0(\omega) \mathbf{F}(\mathbf{r},\omega). 
\end{equation}
Here we have introduced the dielectric constant 
\begin{equation}
    \epsilon(\omega) = 1 + \eta^2 \chi_0(\omega). 
\end{equation}
We now can use this to eliminate the electric field formally as 
\begin{equation}
    \mathbf{E}(\mathbf{r},\omega) =  \int d^3 r' \mathbb{\hat{G}}^R(\mathbf{r},\mathbf{r}';\omega)\left[ -\eta \omega^2 \chi_0(\omega) \mathbf{F}(\mathbf{r}',\omega) \right] 
\end{equation}
where 
\begin{equation}
     \mathbb{\hat{G}}^R(\mathbf{r},\mathbf{r}';\omega) =   \left[ \omega^2 \epsilon(\omega)\mathds{1} - \nabla \times \nabla \times \right]^{-1}  .
\end{equation}
We can evaluate this Greens's function in a manner of our choosing; in a bulk system it makes sense to use momentum space functions. 

The full response of the phonons, dressed by the photons,
\begin{equation}
    \mathbf{Q}(\mathbf{r},\omega) = \int d^3 r' \mathbb{\hat{D}}^R(\mathbf{r},\mathbf{r}';\omega) \mathbf{F}(\mathbf{r}',\omega),
\end{equation}
is given as a function of the photon Green's function:
\begin{equation}
    \mathbb{\hat{D}}^R(\mathbf{r},\mathbf{r}';\omega)  = \chi_0(\omega) \left[ \delta^3(\mathbf{r}-\mathbf{r}')\mathds{1}  - \eta^2\omega^2  \chi_0(\omega) \mathbb{\hat{G}}^R(\mathbf{r},\mathbf{r}';\omega) \right] .
\end{equation}
In the absence of phonon dispersion the first term is completely local, and exhibits a resonance at the bare TO mode frequency, while the second term involves dispersion of the phonon polaritons in the system, as it involves $\epsilon(\omega)$, and therefore is sensitive to the cavity geometry. 

We note we can write this in an elegant way by using  
\begin{equation}
    -\eta^2 \left( \chi_0(\omega) \right)^2 = \frac{\delta \epsilon(\omega)}{\delta\Omega_T^2 } 
\end{equation}
to obtain 
\begin{equation}
\begin{split}    \mathbb{\hat{D}}^R(\mathbf{r},\mathbf{r}';\omega)  =& \chi_0(\omega) \delta^3(\mathbf{r}-\mathbf{r}')\mathds{1}  + \\
&\omega^2\frac{\delta \epsilon(\omega)}{\delta \Omega_T^2 } \mathbb{\hat{G}}^R(\mathbf{r},\mathbf{r}';\omega) .
\end{split}
\label{eqn:DressedProp}
\end{equation}

\subsection{Matsubara Formalism}
\label{sub:matsubara}
The result presented is derived using a real-time linear-response formalism, which is the most transparent presentation.
In Appendix~\ref{app:matsubara}, we equivalently derive this result using the equilibrium Matsubara frequency representation, as well as yet a third way of deriving this result based on a variational procedure for the Matsubara free-energy functional in Appendix~\ref{app:variational}.

The Matsubara frequency representation is particularly useful since it allows for a more efficient implementation of the result in terms of a well-behaved, convergent sum over Matsubara frequencies. 
For more details, we refer the reader to the relevant Appendices.
However, we present the end formula here as it is relevant for the results to follow.

The dielectric function is analytically continued to bosonic Matsubara frequencies $\omega \to i\omega_m = 2\pi i m T$ as 
\begin{equation}
    \epsilon(i\omega_m; \mathbf{r}) = 1 + \frac{\eta^2}{\omega_m^2 + \Omega_T^2(\mathbf{r})},
\end{equation}
allowing for a locally varying TO mode frequency. 
Likewise, the unscreened phonon propagator becomes 
\begin{equation}
    \mathscr{\hat{D}}_0(\mathbf{r},\mathbf{r}';\omega_m)= \frac{1}{\omega_m^2 + \Omega_T^2(\mathbf{r})}\delta^3(\mathbf{r}-\mathbf{r}'),
\end{equation}
and the photon propagator becomes 
\begin{equation}
    \mathscr{\hat{G}}(\mathbf{r},\mathbf{r}';\omega_m)= \left[ \omega_m^2 \epsilon(i\omega_m,\mathbf{r}) + \nabla \times \nabla\times \right]^{-1}.
\end{equation}
The local phonon fluctuations as a function of temperature are calculated by combining equations~(\ref{eqn:FDT}),(\ref{eqn:DressedProp}), which is conveniently expressed in terms of a Matsubara sum as
\begin{multline}
\label{eqn:matsubara-QQ}
 \langle \mathbf{Q}^2(\mathbf{r},t)\rangle = T \sum_{\omega_m} \\
 \tr\left[\mathscr{\hat{D}}_0(\mathbf{r},\mathbf{r}; \omega_m) + \omega_m^2 \frac{\partial \epsilon(i\omega_m,\mathbf{r})}{\partial\Omega_T^2} \mathscr{\hat{G}}(\mathbf{r},\mathbf{r};\omega_m)\right]
\end{multline}
Note that 
\begin{equation}
    \frac{\partial \epsilon(i\omega_m,\mathbf{r})}{\partial\Omega_T^2} = -  \frac{\eta^2}{(\omega_m^2 + \Omega_T^2(\mathbf{r}))^2}
\end{equation}
is negative semidefinite, while the other terms in the equation are positive semidefinite.
This leads to a natural conclusion that the electric field actually has the effect of {\bf reducing} phonon fluctuations, since it subtracts spectral weight away from the otherwise unscreened phonons. 
We also see that the result can be implemented by a sum over terms which are non-singular, greatly aiding the numerical evaluation of this expression. 
This is somewhat similar to the recent approach of Ref.~\cite{Lenk.2022}, though extended to the multimode formalism.
Nevertheless, the concept of a $1/N$ expansion would be useful to apply in this context as well. 

We can then ultimately express quantities in terms of the effective frequency renormalization as compared to the bulk value
\begin{equation}
    \delta \Omega_T^2(\mathbf{r}) = \lambda \left[ \langle \mathbf{Q}^2(\mathbf{r})\rangle - \langle \mathbf{Q}^2\rangle_{\rm bulk}  \right].
\end{equation}
For each configuration, this is done by extracting the bulk value as the value computed as system size tends to infinity. 
We can then visualize the predicted spatial deviations of the frequency away from the bulk value. 
We now roughly estimate the size of the coupling $\lambda$, the momentum space cutoff $\Lambda$, and other relevant parameters.

\subsection{Parameters}
\label{sub:parameters}
Following Ref.~\cite{Katayama.2012}, we first extract the value of the anharmonic potential $U^{(2)}$ as a function of the ionic displacement coordinate $\mathbf{u}$, for the case of SrTiO$_3$, estimated from spectroscopy to give $\nu_T \sim 35\si{\THz^2\per\pico\meter^2}$.
This is obtained from nonlinear terahertz spectroscopy by measuring $\Omega_{T}^2(\mathbf{u}) \sim \Omega_T^2 + \nu_T \mathbf{u}^2 + O(\mathbf{u}^4)$.
However, this is not directly related to the long-wavelength order parameter~\footnote{This is similar to the case of Ginzburg-Landau theory where, in general the microscopic parameters such as density-of-states and quasiparticle gap are not directly related to the long-wavelength order parameter. Rather, only certain ratios of coupling constants are fixed in terms of microscopic parameters.}, which in general involves a complicated coarse-graining of the microscopic degrees of freedom.
By dimensional analysis, we see that the long-wavelength coupling constant can be related to a microscopic parameter through $\lambda \sim \nu_T V_{\rm eff}/M_{\rm eff}$ where $M_{\rm eff}$ is an effective mass scale and $V_{\rm eff}$ is an effective volume scale.
For the purposes of our crude model, we use a single effective mass to characterize the mode, which we take to be the titanium ionic mass $M_{\rm eff} \sim M_{\rm Ti} \sim 50 \textrm{amu,}$

Finally, in the spirit of a real-space renormalization group procedure, we anticipate that the effective volume $V_{\rm eff}$ which appears in the relation between the order-parameter and the microscopic degrees of freedom should be related to the momentum space cutoff we place on our model, $\Lambda\sim 1/a$ with $a$ the size of the coarse-grained ``blocks."
As a result, we estimate that $V_{\rm eff}\sim 1/\Lambda^3 \sim a^3$. 
We thus identify the long-wavelength coupling as $\lambda = a^3 \nu_T /M_{\rm Ti}$.
This leads to the cutoff-dependent estimate of 
\begin{equation}
    \lambda = 32\times 10^4 \si{\THz^3} a^3,
\end{equation}
which we use in this work.

Finally, by ignoring the gradient terms $\sim \nabla \mathbf{Q}$, our model treats the phonon fluctuations as purely local.
In reality, we expect this approximation to breakdown below a length scale $a$, roughly related to the correlation length of the ferroelectric order parameter $\mathbf{Q}(\mathbf{r})$.
On the one hand, near the critical point this length scale may become large as correlations develop across longer length scales.
One the other hand, treating physics at a length scale $l < a$ requires a more complicated theory than the one we develop here. 
In this work we are interested in macroscopic physics, with samples of order $L\sim 50\si{\micro\meter}$ or larger; as a result, we will consider a cutoff of $a = 5 \si{\micro\meter}$.

In the future it would be particularly interesting to consider going even closer to the critical point, so that $a \sim L$, in which case electrodynamic control may be even more important.
This is because the photon dispersion is much faster than the phonon, and thus the region in reciprocal space amenable to cavity control is roughly $q_{\rm QED}^3 \sim (1\si{\tera\hertz}/c)^3 \sim (.3\si{\milli\meter})^{-3}$, which is to be compared against the volume of the Brillouin zone which exhibits fluctuations, which goes roughly as $1/a^3$.
As such, we expect the physics we consider to be most important materials which are close to the ferroelectric instability, such that $a^{-1}\sim q_{\rm QED}$. 
However, in this regime a more elaborate model which considers the role of phonon dispersion and spatial gradients is required.
Evidently, a more elaborate scaling theory of the transition is needed in the future.
We elaborate on this slightly later, in Sec.~\ref{sec:conclusion}.



Finally, we must fix the phonon TO and LO mode frequencies. 
For the bulk TO frequency $\Omega_T$ and LO-TO splitting $\sim \eta$, we take $\Omega_T = .5$THz to emulate the ferroelectric soft mode, and $\eta = 10$THz, such that $\epsilon(0)\sim \eta^2/\Omega_T^2 \sim 400$ is large, as is the case for many incipient ferroelectrics. 
We also ignore the temperature dependence of these parameters for simplicity, though this could be accounted for more careful if we were modeling a specific material. 

Finally, for numerical purposes we take a Matsubara frequency cutoff of $\omega_c = 40$THz, well above all other frequency scales; this is not expected to be an important parameter, provided it is large enough. 
In order to make the calculations simpler and more transparent we also relax the self-consistency demand on the TO frequency, and instead simply evaluate the perturbative correction to the bulk value.
In principle, this should be addressed though if the shift is small it is likely to be qualitatively correct.

\section{Homogeneous System}
\label{sec:homogeneous}
Before proceeding on to our main calculation, we briefly outline how the calculation works in the case of a system with full translational symmetry.
In this case, momentum is a good quantum number in all directions. 
We can then go to the plane-wave basis. 

Using 
\begin{equation}
\nabla\times\nabla \to -\mathbf{q}\times\mathbf{q}\times =  \mathbf{q}^2\mathds{1} - \mathbf{q} \otimes \mathbf{q}  
\end{equation}
we find the electromagnetic correlation function splits into two decoupled subspaces: the transverse modes and longitudinal modes, with 
\begin{multline}
    \omega_m^2 \mathscr{\hat{G}}(\omega_m,\mathbf{q}) = \\
    \frac{\mathbf{q}\otimes\mathbf{q}}{\mathbf{q}^2}\frac{1}{\epsilon(i\omega_m) }  + \left(\mathds{1} - \frac{\mathbf{q}\otimes\mathbf{q}}{\mathbf{q}^2 } \right)\frac{\omega_m^2}{\omega^2_m\epsilon(i\omega_m) + \mathbf{q}^2 }  . 
\end{multline}
We easily recognize the first term as the dynamical Coulomb portion of the electric field, while the second term comes from the transverse photon modes. 
The unscreened phonon propagator is trivial in this case, with 
\begin{equation}
    \mathscr{\hat{D}}_0(\omega_m,\mathbf{q}) = \frac{1}{\omega_m^2 + \Omega_T^2} . 
\end{equation}

We find for the overall result, with UV cutoff on momentum $\Lambda$,
\begin{multline}
    \langle\mathbf{Q}^2\rangle = T \sum_{\omega_m}\int_{|{\bf q}| < \Lambda} \\
    \left[ \frac{3}{\omega_m^2 + \Omega_T^2} + \frac{\partial \epsilon(i\omega_m)}{\partial\Omega_T^2}\left( \frac{1}{\epsilon(i\omega_m)} +  \frac{2 \omega_m^2}{\omega_m^2 \epsilon(i\omega_m) + \mathbf{q}^2 } \right) \right].
\end{multline}
This simplifies into one longitudinal mode and $d-1$ transverse modes (with $d$ the spatial dimensionality), with the longitudinal modes contributing (per momentum space mode)
\begin{multline}
    T \sum_{\omega_m}\left[ \frac{1}{\omega_m^2 + \Omega_T^2} + \frac{\partial \epsilon(i\omega_m)}{\partial\Omega_T^2}\frac{1}{\epsilon(i\omega_m)} \right] \\
    = \frac{\coth(\beta \Omega_L/2)}{2\Omega_L}  .
\end{multline}
The momentum space integral will then simply yield a factor of $\Lambda^3/6\pi^2$, with the cubic divergence due to the dispersionless nature of the model. 

The transverse modes are more difficult to evaluate.
Evaluating the Matsubara sums and performing the analytical continuation to real frequencies reveals that the overall result including the longitudinal and two transverse modes is  
\begin{multline}
    \langle \mathbf{Q}^2\rangle = \int_{|{\bf q}|<\Lambda} \bigg[\frac{\coth(\beta \Omega_L/2)}{2\Omega_L } \\
    + 2\times \frac{\coth(\beta \Omega_{+,q}/2)}{2\Omega_{+,q} }\frac{\Omega_{+,q}^2 - q^2}{\Omega_{+,q}^2-\Omega_{-,q}^2} \\
    + 2\times \frac{\coth(\beta \Omega_{-,q}/2)}{2\Omega_{-,q} }\frac{\Omega_{-,q}^2 - q^2}{\Omega_{-,q}^2-\Omega_{+,q}^2}\bigg].
\end{multline}
The dispersion of the upper polariton branch ($\Omega_{+,q}$) and lower polariton branch ($\Omega_{-,q}$) are found to be
\begin{equation}
    \Omega_{\pm,q} = \sqrt{ \frac{\Omega_L^2 + q^2}{2} \pm \sqrt{ \left( \frac{\Omega_L^2 + q^2}{2} \right)^2 - \Omega_T^2 q^2 }}.
\end{equation}
This is in accordance with the result one would expect from Bogoliubov transformation and diagonalizing the quasiparticle Hamiltonian. 


\subsection{High-Temperature Limit}
\label{sub:high-temp}

We also can look at the temperature dependence of this effect, and in particular at high-temperatures we obtain, after simplifying terms
\begin{equation}
    \langle \mathbf{Q}^2\rangle = T \int_{|{\bf q}|<\Lambda} \bigg[\frac{1}{\Omega_L^2 } + \frac{2}{\Omega_T^2} \bigg].
\end{equation}
This is exact and independent of momentum $\mathbf{q}$.

In this case we see the electromagnetic field only serves to split the longitudinal modes away from the transverse modes, which then essentially decouple form the electromagnetic field fluctuations.
All the non-trivial electrodynamic effects end up vanishing as $1/T$, which can be interpreted as them being a true manifestation of quantum effects. 
This is evident from examining the relevant Matsubara sums, which end up going as $\omega_m^2/q^2$ at low frequencies, and therefore vanish in the static $\omega_m = 0$ limit.
The finite Matsubara frequencies are in turn arising from dynamical quantum fluctuations of the field, which are ultimately the source of the polaritonic splittings. 

\begin{figure}
    \centering
    \includegraphics[width=\linewidth]{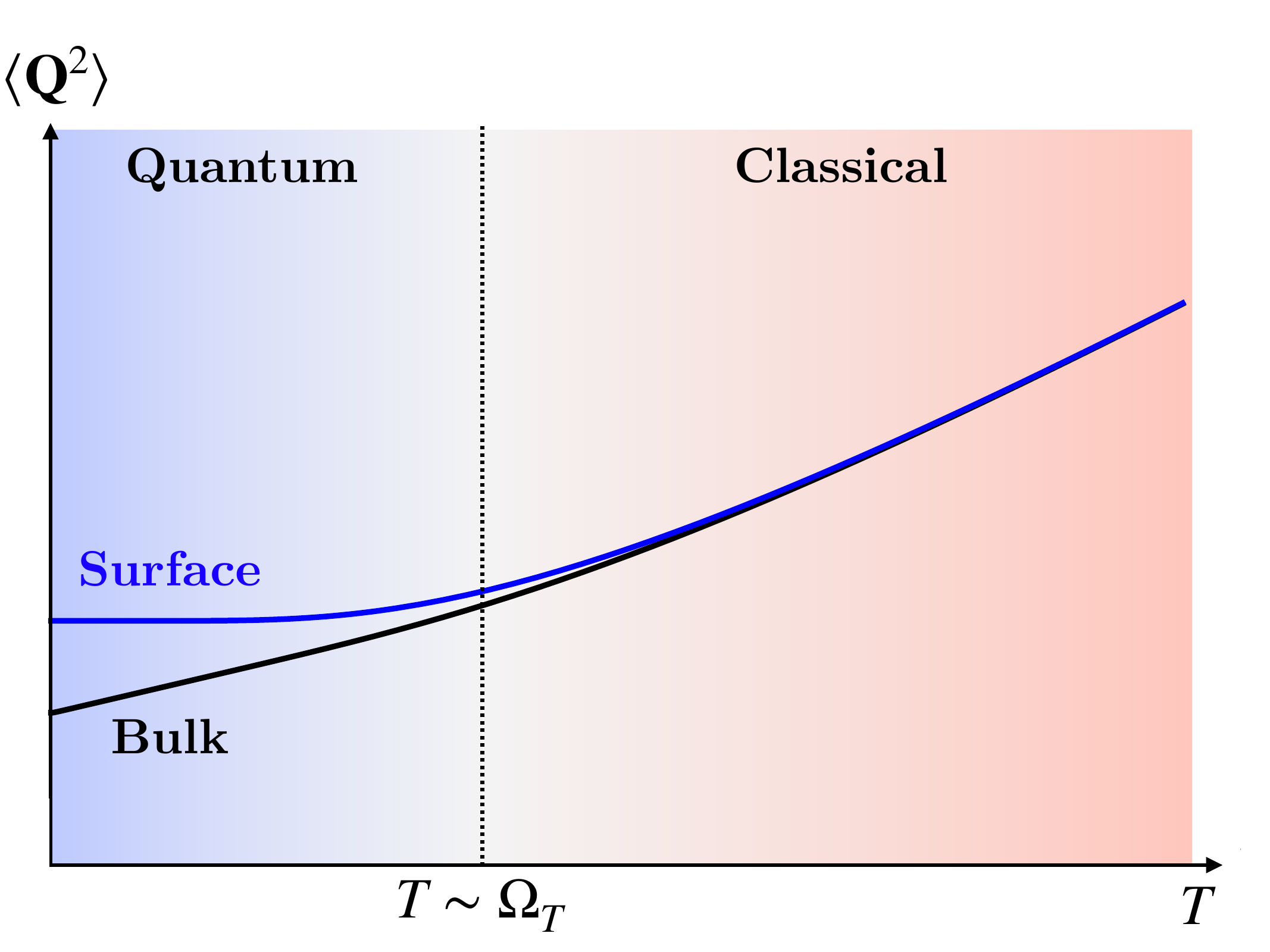}
    \caption{Schematic depiction of temperature dependence of phonon fluctuations $\langle \mathbf{Q}^2\rangle $ as a function of temperature in the bulk and near the surface of the sample-cavity boundary. 
    At high temperatures the fluctuations are independent of electrodynamic details and therefore are ignorant of the proximity to the surface.
    At lower temperatures, the electrodynamic contribution becomes important and leads to a more pronounced fluctuations near the surface as compared to the bulk.}
    \label{fig:surf-bulk}
\end{figure}

We emphasize that this does not mean that the fluctuations of $\langle \mathbf{Q}^2\rangle$ diminish with increasing temperature, but rather that the contribution from the electromagnetic field falls off.
As a result, if one performs a high-temperature expansion of $\langle\mathbf{Q}^2\rangle$ in the bulk one would get $\langle \mathbf{Q}^2 \rangle_{\rm bulk} \sim a_{\rm bulk} T + b_{\rm bulk}/T+...$, with the leading term the classical equipartition result and the subleading terms coming from the high-temperature expansion of $\coth(\omega\beta/2)$ (which is odd, so the series has only odd powers of temperature).
Performing the same expansion for the fluctuations near the surface (which as we will show, differs in that it doesn't couple to the transverse photons due to boundary conditions) we get $\langle \mathbf{Q}^2 \rangle_{\rm surf} \sim a_{\rm surf} T + b_{\rm surf}/T+...$ and find that $a_{\rm bulk } = a_{\rm surf}$, so that the effects of the boundary conditions are subleading in $T$ and due to truly quantum effects.
This is illustrated in Fig.~\ref{fig:surf-bulk} schematically, which shows that at high temperatures the classical result dominates and only at low temperatures is the interaction with photons relevant. 

This is reminiscent of the Bohr-Van Leeuwen theorem in classical mechanics, and can be heuristically understood as a consequence of the current and displacement being canonically conjugate for the phonon.
This is explored more technically in detail Appendix~\ref{app:high-temperature}; here we provide a simple heuristic.
Essentially, photons don't actually couple directly to the phonon displacement $\mathbf{Q}$, but rather couple to the displacement current $\mathbf{J} \sim \partial_t\mathbf{Q}$ associated to transverse oscillations in the charge distribution. 
This current is canonically conjugate to the object of interest, $\mathbf{Q}$, via $[\mathbf{Q}(\mathbf{r}),\mathbf{J}(\mathbf{r}')] \propto i\hbar \delta^3(\mathbf{r}-\mathbf{r}') $ and therefore in the quantum system (at low temperatures) they are not independently fluctuating.
Thus, modifying the fluctuations of the current $\mathbf{J}$ can in turn induce changes in the fluctuations of $\mathbf{Q}$. 
However, this coupling is purely due to the canonical commutation relations between the two fields, and therefore at high-temperatures (i.e. in the classical limit), the two variables become independently fluctuating, just as $q$ and $p$ are independent in a classical system.
Therefore, the photon decouples from the fluctuations of $\mathbf{Q}$ since this is now independent from the current.
We also see that this is not the case for the longitudinal part, which instead directly couples the electrostatic potential to the induced phonon charge $\rho \sim \nabla \cdot\mathbf{Q}$, and indeed the LO-TO splitting due to this interaction remains unaffected in the high-temperature classical limit. 

To summarize, we see directly from this calculation that the coupling to the electric field reduces the fluctuations of the phonons, both by dressing the eigenfrequencies, and also by reducing the projection of the quasiparticle wavefunction onto the phononic subsystem.
Therefore, we uncover a counter-intuitive heuristic that coupling to the quantum electromagnetic field actually facilitates ferroelectric order, by virtue of reducing the fluctuation-induced shift of phonon frequency.
In the following section we will apply this formalism to the case of an inhomogeneous slab structure, and demonstrate how this is potentially visible in terms of a local shift in the phonon mode frequency.

\section{Planar Geometry}
\label{sec:fabry-perot}
We now evaluate the correlation functions needed to compute the frequency shift.
In particular, we focus on the electromagnetic Green's function since the phonon correlation function is local and trivial to solve.
More over, this term will essentially be a bulk background contribution, and in this work we are focused on the contribution from the degrees of freedom we can change by the boundary conditions (the photons).

In the planar geometry we can utilize in-plane translation symmetry to reduce the problem to a one-dimensional differential equation in terms of the spatial coordinate $z$.
It turns out this can still be solved analytically utilizing the transfer matrix method, at least in the case of a homogeneous dielectric constant, as was originally done a long time ago for the purposes of calculating Casimir-Polder forces (which turns out to be a closely related problem)~\cite{Lifschitz.1956,Dzyaloshinskii.1961,Kenneth.2006}.
We can take the momentum to lie in the $x$-direction, such that the Green's function for the gauge potential $\mathscr{G}$ reads 
\begin{equation}
    \left[ \omega_m^2 \epsilon(i\omega_m)\mathds{1} + \begin{pmatrix}
    -\partial_z^2 & 0 & i q \partial_z \\
    0 & q^2 - \partial_z^2 & 0 \\
    i q \partial_z & 0 & q^2 \\
    \end{pmatrix}\right]\mathscr{G}(z,z') = \mathds{1}\delta(z-z') .
\end{equation}
This system is depicted schematically in Fig.~\ref{fig:schematic}, illustrating the metal-paraelectric-metal geometry. 
In this work, we take the physically reasonable limit of infinite plasma frequency in the metal, such that the metal mirrors can be modeled by Dirichlet boundary conditions on the tangential components of $\mathbf{E}$ and normal component of $\bf B$. 
We take the cavity plates to be located at $z = \pm L/2$ such that $L$ is the total size of the cavity, and also the full extent of the paraelectric is all the way up to the boundaries.

This is solved in detail in Appendix~\ref{app:solution}, we will only present the final result here. 
We find that the trace of the Green's function evaluated at coincident spatial points $\tr \mathscr{G}(z,z)$ is given in closed form as 
\begin{multline}
    \tr \mathscr{G}(z,z) = \frac{\sinh\kappa(L/2 - z)\sinh\kappa (L/2 + z)}{2\kappa \sinh \kappa L} \\
    + \frac{1}{\omega_m^2 \epsilon(i\omega_m)}\delta(0).
\end{multline}
Here $\kappa = \sqrt{\omega_m^2 \epsilon(i\omega_m) + q^2}$ governs the length-scale for the recovery to the bulk value for a given frequency and in-plane momentum.
We see that this involves the divergent quantity $\delta(0)$, which is understood as $\lim_{z'\to z}\delta(z-z')$.
This quantity is in fact independent of system-size and geometry and thus ends up getting renormalized away by the counter-term $\Omega_0^2$.
In particular, we only compare this result to result obtained for $L\to \infty$ (with $z$ finite), which ultimately gives the closed-form formula for the renormalization of the phonon-frequency shift due to the cavity, as a function of position, as 
\begin{widetext}
\begin{equation}
\label{eqn:master-formula}
    \left(\Delta \Omega_T(z)\right)_{\rm cav}^2 = \lambda T \sum_{\omega_m}\int^\Lambda \frac{d^2 q_{\parallel}}{(2\pi)^2}  \frac{\partial \epsilon(i\omega_m)}{\partial \Omega_T^2} \frac{\omega_m^2}{2\kappa}\left[ \frac{\sinh\kappa(L/2 - z)\sinh\kappa (L/2 + z)}{\sinh\kappa L} - \frac12 \right].
\end{equation}
\end{widetext}
We note this expression does still depend on the cutoff $\Lambda = \pi/a$. 
While this is much easier to evaluate than the full Green's function, it is still not completely trivial due to the subtle behavior of the integrand at $\omega_m \to 0$, which governs to the high-temperature limit of the effect. 
For all non-zero Matsubara frequencies, this is easily evaluated by numerically summing, for $\omega_m = 0$ there is an issue about how the limit of zero frequency is taken; on the one hand, the numerator involves a power of $\omega_m^2$ which then vanishes quadratically at small frequency.
On the other hand, the denominator involves $\kappa \sinh \kappa L$, which for $q_\parallel \to 0 $ also vanishes quadratically as $|\omega_m| \sinh |\omega_m|\sqrt{\epsilon(0)} L $.

Careful examination of the limit reveals that the numerator ends up winning and thus, we are to discard altogether the zeroth frequency contribution.
This is again a manifestation of the quantum mechanical origin of the effect, as explained in Sec.~\ref{sub:high-temp} and shown in greater detail in the Appendix~\ref{app:high-temperature}. 
We are now able to efficiently evaluate this effect in order to obtain not just the phonon frequency shift but its entire spatial dependence. 
We also comment that this expression is clearly manifestly positive, owing to the inequality $\sinh \kappa (L/2 - z) \sinh \kappa (L/2 + z)/\sinh \kappa L \leq \frac12 $.
As a result, we find the frequency shift is always positive, with $\Delta \Omega_T^2(z) \geq 0$, such that the cavity in fact {\em suppresses} the ferroelectric order. 
We will comment on this more later. 

\begin{figure}
    \centering
    \includegraphics[width=\linewidth]{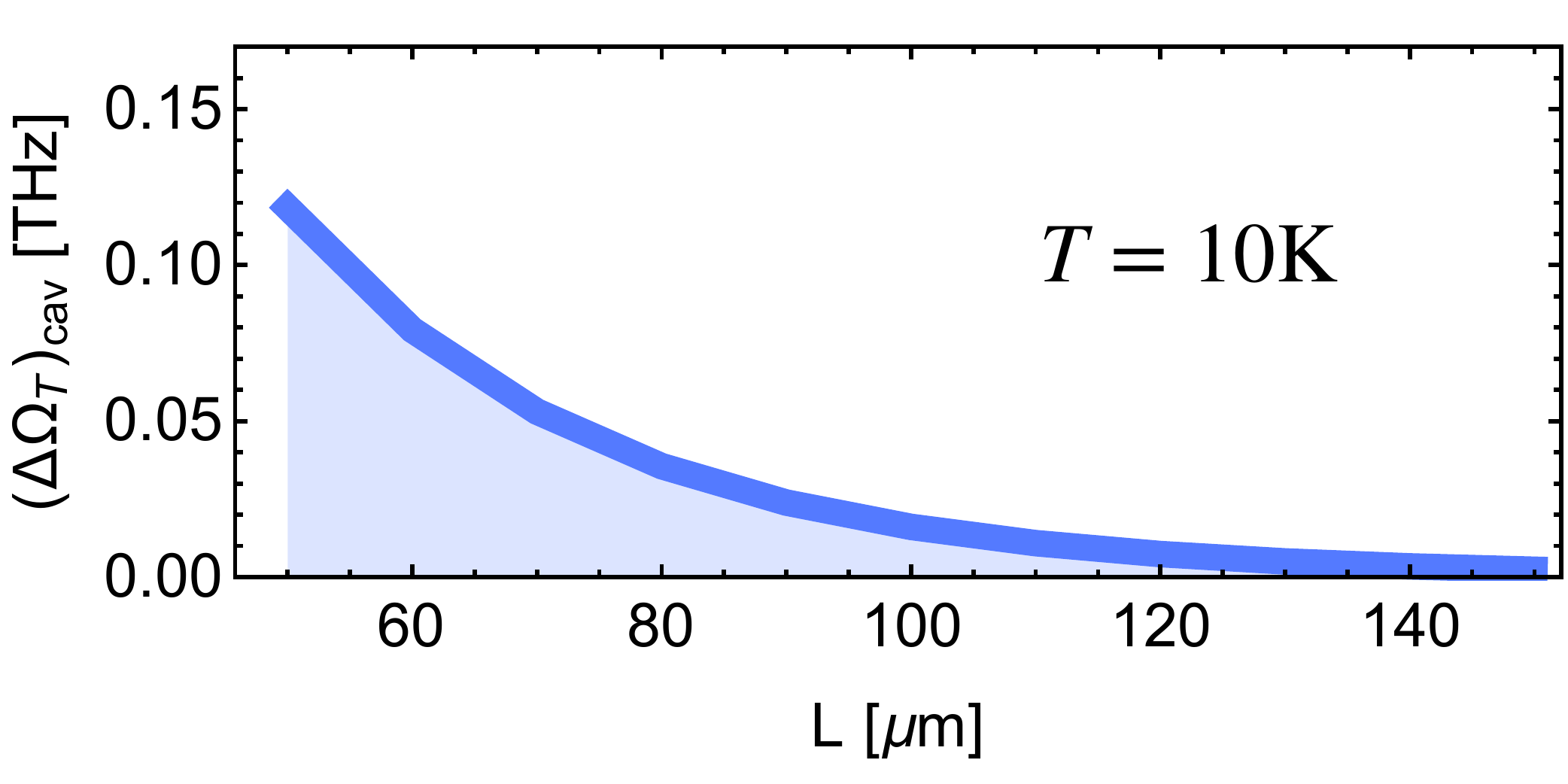}
    \caption{First order correction to phonon frequency due to cavity boundary conditions at $z = \pm L/2$.
    We here fix the temperature to $T = 10K$, which is lower than the coherence temperature for the phonon frequency of $\Omega_T = .5$THz (we note the coherence frequency involves a factor $2\pi$ so that $ 2\pi T \lesssim \Omega_T$).
    We explicitly note that this is the correction to the phonon frequency due to the cavity and in particular, this blue shifts at lower temperature, unlike the typical phonon anharmonicity which blue shifts at higher temperatures. 
    The blue shift due to the anharmonicity is renormalized away in this treatment; here we only plot the additional shift observed between the bulk and finite systems at the same temperature.}
    \label{fig:L-dependence}
\end{figure}

In particular, it is interesting to analyze the dependence of the renormalized frequnecy on system size $L$ and temperature $T$ (we consider the value in the midpoint of the cavity). 
This is shown in Fig.~\ref{fig:L-dependence}.
We indeed see that for ``bulk systems" with large system size $L$, there is no effect due to the boundary conditions (which is expected on grounds of locality).
For smaller systems however, the typical phonon frequency strongly blue shifts as the surface effects set in.
For very small systems, the entire system is essentially ``surface" and thus we see the first characteristic prediction which is a significant blue-shifting of the soft-mode frequency for thin samples.

\begin{figure}
    \centering
    \includegraphics[width=\linewidth]{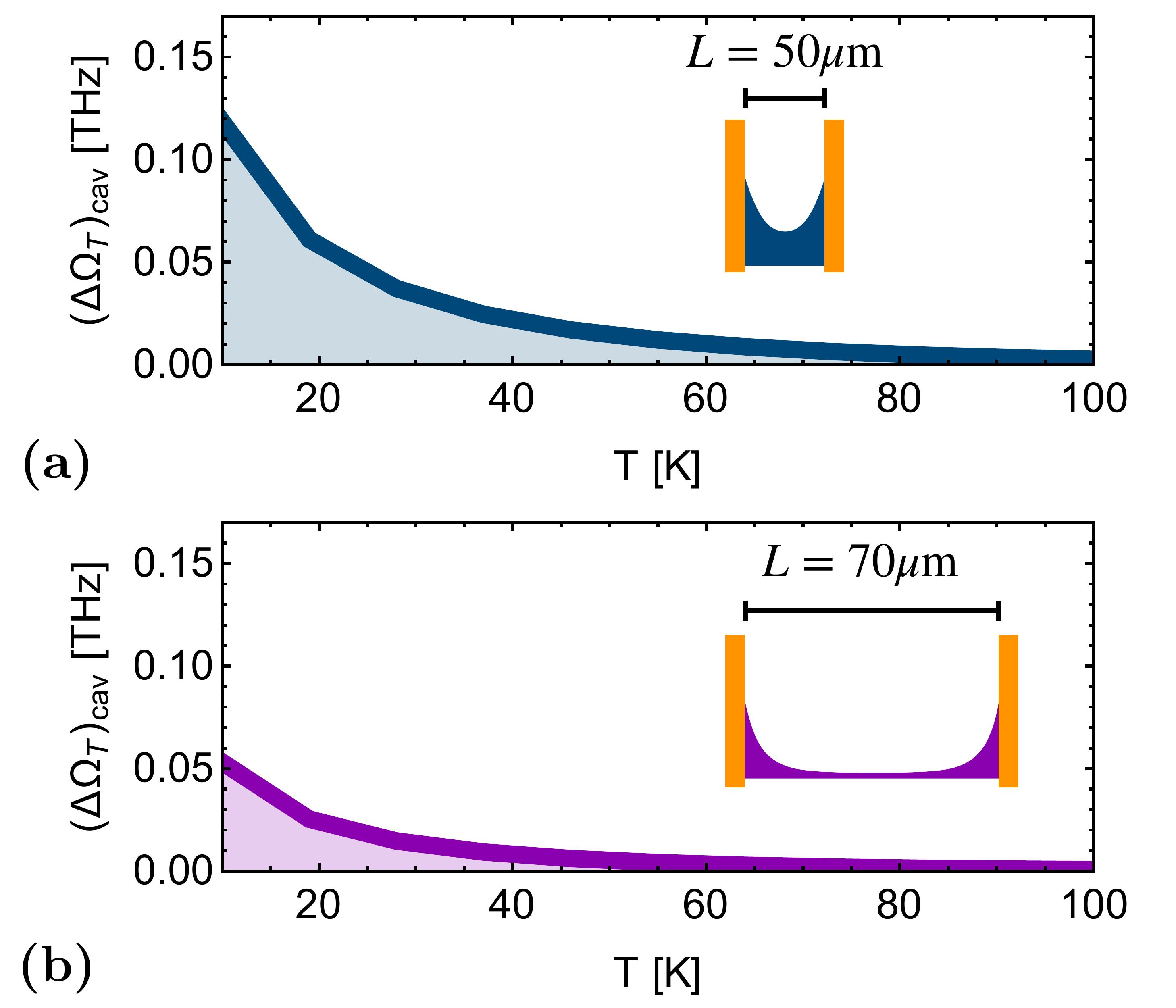}
    \caption{First order correction to phonon frequency due to the cavity as a function of temperature $T$.
    Here we consider two cases: a small system with $L = 50\si{\micro\meter}$ (a), and a larger system with $L=70\si{\micro\meter}$ (b).
    In the inset we depict schematic profiles of the renormalized frequency, illustrating how the effect is larger for smaller systems since the surface effects are still dominant, whereas in a larger system the frequency converges to the bulk value. }
    \label{fig:T-dependence}
\end{figure}

We next confirm the temperature dependence of this effect; namely, that at high-temperatures any signature of the cavity should disappear. 
This is seen explicitly in Fig.~\ref{fig:T-dependence}.
Indeed we see that in all system sizes, the frequency shift vanishes at high temperature irrespective of the system size $L$.
At lower temperatures, the phonon frequency shift sets in, but for larger systems the effect is small since it ultimately must recover to the bulk as $L\to \infty$.
For smaller systems however, the shift may become sizeable at the cavity midpoint.

\begin{figure}
    \centering
    \includegraphics[width=\linewidth]{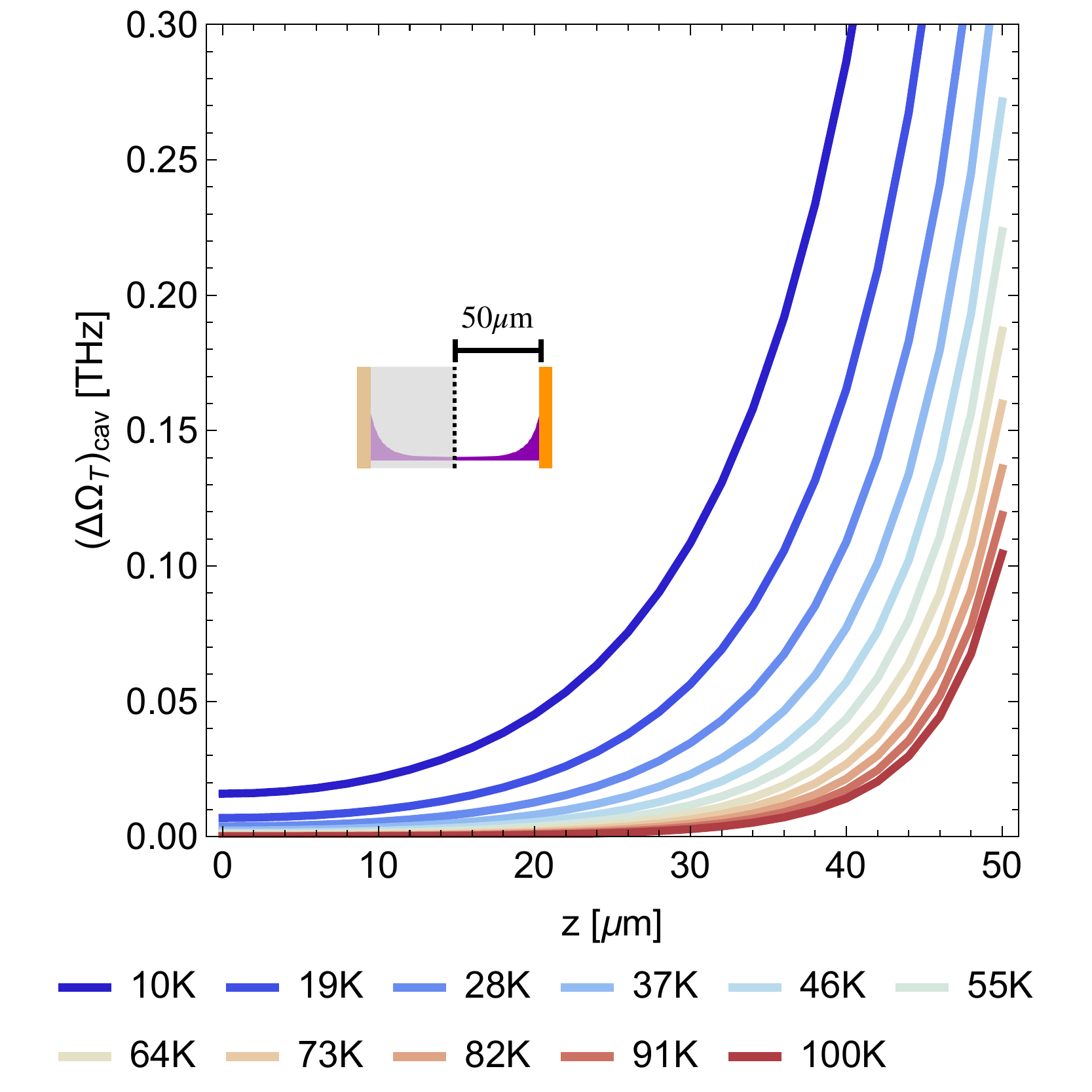}
    \caption{Perturbatively renormalized phonon frequency $\Omega_T(z)$ including the electrodynamic correction due to cavity boundary conditions as a function of spatial coordinate $z$.
    System size is fized at $L = 100\si{\micro\meter}$, with coordinate $z$ referenced from the midpoint, as shown in the inset.
    Temperature varies from high temperature of $T = 100$K (with little effect) down to low tmeperature of $T = 10$K, with a pronounced blue-shift in the local phonon frequency occurring at the boundary. }
    \label{fig:z-dependence}
\end{figure}

These results are best understood by simply looking at the spatial profile for the renormalized frequency $\Omega_T^2(z)$, depicted in Fig.~\ref{fig:z-dependence}.
For a cavity size of $100\si{\micro\meter}$ we see that the phonon frequency significantly blue-shifts near the boundary at low temperatures, while it remains essentially equal to the bulk value at high temperatures and in the center of the cavity $z \sim 0$).
This is inline with the previous arguments we have given, namely that the cavity renormalization is (i) a purely quantum effect setting in once $2\pi T \sim \Omega_T$, and (ii) that it is a surface effect in response to the boundary conditions at $z =\pm L/2$. 
We also always see that the frequency blue-shifts\textemdash this is in some tension with a number of previous theoretical investigations~\cite{Ashida.2020,Latini.2021,Lenk.2022,Lenk.2022b}, which all seem to find that cavities tend to enhance ferroelectric order. 
We now reconcile our calculation with these previous studies.

We believe the origin of this tension is in the way that the UV cutoff on the electromagnetic fluctuations is imposed, and ultimately scaled to the continuum limit (or not). 
In particular, in a real physical system (with an appropriate UV cutoff) the number of electromagnetic modes is proportional to system size, with a finite number of modes (e.g. lattice points) per unit volume, albeit of potentially high frequency and thus far from resonance.
We again emphasize that it is qualitatively important to take into account the multimode nature of the electromagnetic fluctuations~\cite{Amelio.2021,Andolina.2022b}. 

In our new formalism, which up to, and including Eqn.~\eqref{eqn:master-formula} is an analytically exact solution to the problem, the UV cutoff is essentially scaled in this fashion, so that the density of modes per unit volume is constant.
This is also in congruence with known and experimentally verified results on Casimir forces~\cite{Casimir.1948,Lifschitz.1956,Dzyaloshinskii.1961,Kenneth.2006}.
In contrast, previous studies have essentially imposed a cutoff on number of cavity eigenfunctions taken, selecting the lowest $N_c$ modes to include in an eigenmode expansion for the Green's function $\mathscr{G}$ regardless of system size.
This assumption leads to the unphysical result that number of electromagnetic modes per unit volume $\sim N_c/L^3$ is scaling to zero for a bulk system!
It turns out that, unfortunately this has exactly the opposite behavior as what we find using the more complicated analytical solution~\footnote{The authors would like to especially acknowledge private correspondence with Yuto Ashida on this point, who patiently and carefully worked through this subtlety with us, and who ultimately helped uncover the origin of the discrepancies. Discussions with Ata\c{c} {\. I}mamo{\u g}lu were also valuable in this regard.}.

While our approach draws upon parallels with Casimir forces, it is however investigating a distinct phenomenon and thus it still warrants experimental confirmation.
We now briefly discuss the prospects for this in the next section.

\section{Discussion}
\label{sec:conclusion}

As we saw in the previous sections, the effect of the cavity ends up being largely confined to surface of the material, with the net result being a blue shift of the soft phonon mode near the boundaries. 
This would result in an apparent diminishing of the dielectric constant $\varepsilon(\omega = 0)\sim \eta^2/\Omega_T^2$ for a small sample in comparison to the bulk value.
In fact, this effect has already been observed and known for quite some time as the ``dead-layer" effect~\cite{Zhou.1997}.
Canonically, this effect was found in exactly this sort of system, comprised of a thin layer of SrTiO$_3$ sandwiched between two electrodes to form a capacitor~\cite{Sirenko.2000}. 
Historically, the dead-layer effect has been attributed largely to strain effects~\cite{Katayama.2008} between the two interfaces which also acts to blue-shift the soft mode, however it seems the matter has not been entirely settled~\cite{Yang.2016}.
It is quite interesting then in the light of this new work and interpretation to re-open the investigation and determine if any effects due to quantum electrodynamics may be relevant.
To this end, experiments may want to try experiments with different metallic interfaces that induce varying levels of strain, to see how sensitive the dead-layer is to the level of strain, as opposed to the dielectric environment.
Indeed, experiments investigating the interplay between electromagnetic response and the surrounding dielectric environment~\cite{Welakuh.2022} have already shown promising results~\cite{Khatoniar.2022,Kumah.2014,Mannebach.2017}.
Theoretically, this would also warrant further calculations which have open (air) boundary conditions for the SrTiO$_3$ layer, rather than the metallic interface conditions.

Our results also point to another interesting connection which ties our results to studies of the Casimir force~\cite{Casimir.1948,Dzyaloshinskii.1961,Kenneth.2006}.
This is most apparent if one returns to Eqn.~\ref{eqn:master-formula}, which closely mirrors the results for Casimir forces (see, e.g. Ref.~\cite{Kenneth.2006}).
This makes sense since our setup is very similar to the one considered originally by Casimir except in this case the fluctuation force does work on the dielectric constant of the QPE (which is determined self-consistently) rather than the plates of the cavity~\cite{Palova.2009}.
It may be interesting to turn this around and try to utilize Casimir force spectroscopy to probe incipient or critical ferroelectric fluctuations via the effect of fluctuations on radiative forces.

A closely related phenomenon is that of the Van der Waals force, which is also an entropic force due to virtual electromagnetic fluctuations. 
Van der Waals forces typically act between polarizeable molecules and is notably attractive, leading to the total energy being lowered as a result of the coupling to electromagnetic fields~\cite{Casimir.1948,Lifschitz.1956,Dzyaloshinskii.1961}. 
Similarly, the infrared-active phonon field can be thought of as a lattice of polarizeable molecules undergoing virtual polar fluctuations.
The primary difference between the two cases is that in the case of the phonon polaritons the molecules are arranged in a regular lattice, whereas in a fluid the molecules are spatially disordered. 
When placed in a cavity, the metallic boundary serves to screen the electromagnetic field, leading to a net increase in the free-energy as compared to the bulk system, since the Van der Waals forces which get screened out are attractive.

In fact, the extended many-mode nature of this problem is paramount.
This is seen by contrasting our results with the simpler case of a single fluctuating dipole localized near a metallic surface.
In the localized case, the interaction between the single dipole and its image charge (which is a manifestation of the cavity screening) is attractive and reduces the cost of a fluctuating dipolar moment.
However, when we consider an extended distribution of dipoles (as realized by the bulk paraelectric), the result is very different.
In particular, one can check that in the long-wavelength, longitudinally polarized $\mathbf{q} \to 0$ modes remain unaffected by the screening.
To see this, consider an infinite line of longitudinally-polarized dipoles interacting with its image, which is also an infinite line with opposite in-plane component of the dipole moment.
In contrast to the localized case, now the overall interaction is zero since the attractive tip-to-tail interaction for small transverse separations is cancelled by the interaction between distant dipoles, which are essentially tip-to-tip oriented. 
This naturally manifests in our calculations as the LO-TO splitting of the long-wavelength modes remaining unchanged by the boundary conditions.
Transverse modes on the other hand, do interact with the screening effect and this ends up most easily diagnosed by recasting the problem as determining $\epsilon(\omega,\mathbf{r})$ rather than directly determining the dipole-dipole interactions, which become very complicated once quantum dynamics are included.
Our result is also consistent with various recent no-go theorems pertaining to superradiance in cavities~\cite{Andolina.2020,Andolina.2022}.

One may alternatively view this renormalization of the phonon stiffness as a realization of the concept of dynamical localization~\cite{Latini.2021,Ashida.2020,Ashida.2021b}.
This renormalization comes because when the phonon mode oscillates it also linearly couples to the electromagnetic vacuum, and thus the energy and inertia of that mode receive contributions also from the cloud of photons which are attached to the phonon.
By using the cavity to suppress the electromagnetic field, we are essentially decoupling the phonon from its photonic cloud, and this is reflected in the evaluation of the correlation functions for the given geometry.
What is somewhat unanticipated is that this cloud of photons actually makes the phonon mode ``lighter," delocalizing the phonon coordinate. 

We also can understand the purely quantum origin of the effect in this way.
Cavity photons directly couple to the transverse component of the current, $\mathbf{J}\sim \partial_t\mathbf{Q}$ and by using the cavity one can change the radiative renormalization of the current fluctuations.
This can only influence the actual phonon displacement $\langle \mathbf{Q}^2 \rangle$ via the canonical commutation relations, which couple the charge and current together. 
Therefore, observation of this experimentally would truly demonstrate the principle of ``quantum control of quantum materials."

As discussed earlier, in Sec.~\ref{sub:parameters}, we expect our theory to be valid not too close to the critical point, due to the local nature of the approximation we use for the phonon fluctuations. 
We now elaborate slightly on how this may breakdown, and what a more complete theory may look like.  
In particular, the key approximation we made is that the phonon correlation function is purely local, so that $\langle \mathbf{Q}(x) \mathbf{Q}(x')\rangle \sim \delta(\mathbf{r}-\mathbf{r}')$.
This then allowed us to integrate the phonons out in favor of a theory purely in terms of the local dielectric response, $\epsilon(i\omega_m)$.
This approximation is motivated by the observation that since the photon group velocity is much larger than the phonon velocity, all phonon dispersion in the vicinity of the light-cone can be ignored.
Naively, this is a very good approximation, but we do have reason to expect this to breakdown close to the critical point, since at the critical point lattice fluctuations become correlated over much larger length scales than the naive lattice estimate indicates, as one expects from the general theory of critical phenomena.

In order to properly accommodate these correlations, one must introduce phonon dispersion in to the model, with the simplest modification to our current theory given by the imaginary-time Lagrangian (using the same conventions we use in the rest of the paper)
\begin{multline}
        \mathcal{L}_{\rm eff} = \frac12\left(\frac{\partial \mathbf{Q}}{\partial \tau}\right)^2 + \frac12 \Omega_0^2 \mathbf{Q}^2 + \frac{\lambda}{4}(\mathbf{Q}^2)^2 + i \eta \mathbf{Q}\cdot\frac{\partial \mathbf{A}}{\partial \tau} \\
        + \frac12 \left(\frac{\partial \mathbf{A}}{\partial \tau}\right)^2 + \frac12 \left(\nabla \times \mathbf{A}\right)^2 \\
        + \underbrace{\frac12 v_\parallel^2 \left(\nabla\cdot \mathbf{Q} \right)^2 + \frac12 v_\perp^2 \left(\nabla\times \mathbf{Q} \right)^2 }_{\textrm{spatial correlations}}.
\end{multline}
We have emphasize the new terms to be added to accommodate the spatial correlations of the phonon field, which by symmetry in an isotropic medium are characterized by two sound velocities; a transverse mode velocity $v_\perp$ and a longitudinal mode velocity $v_\parallel$.
Unfortunately, including these terms makes the solution technique we employ in this paper more difficult to apply, since it relied on the local nature of the phonon fluctuations. 
In principle, since this modification is only needed near the critical point anyways, it then also makes the mean-field treatment we use for the phonon frequency shift somewhat inapplicable as well.
Instead, in order to proceed we propose employing a renormalization-group (RG) type procedure to handle this theory.
This would then be able to quantitatively assess the degree to which the phonon correlations become important and potentially search for modifications to the critical phenomena due to the long-wavelength modifications due to the cavity geometry, though this is certainly beyond the scope of this paper.
Physically though, it may be interesting to look for modifications to the lattice structure factor and diffuse x-ray scattering due to the presence of the cavity near the critical point, since this may reflect changes to the range of spatial correlations in the incipient lattice distortion. 

We also comment on the relation between what we propose here and recent experiments which seem to indicate optically induced ferroelectricity in SrTiO$_3$ using strong terahertz driving~\cite{Nova.2019,Li.2019ilv}.
It seems likely that in both the present paper and in the terahertz driven experiments~\cite{Nova.2019,Li.2019ilv}, it is important that there be strong precursor fluctuations of a polar soft-mode present in equilibrium.
However, since a conclusive theoretical explanation for the induced ferroelectricity in these systems is still pending, it is difficult to draw a concrete connection.
In particular, owing to the larger magnitudes of electric field accessible in an optically driven material, nonlinear phononic effects may become important~\cite{Kozina.2019} which can complicate the picture by effectively inducing higher order couplings between the electric field and the soft-mode fluctuations.
On the other hand, since the cavity fluctuations are much weaker in amplitude, it is expected that these will more selectively couple to the soft-mode directly, and do so through a predominantly linear polaritonic interaction.
Therefore, it is hard to say with certainty how the mechanism of driven terahertz control is related to the mechanism of cavity control we outline here, other than the fact that both require preexisting fluctuations.

To summarize, we have developed an approach to studying local phonon fluctuations in a QPE interacting with a quantized cavity electromagnetic field.
Rather than making a single-mode approximation or invoking the dipole approximation, we have included a complete continuum of modes which all couple locally to the QPE material.
This allowed us to study the variation in the blue shift of the phonon modes in a spatially resolved way, and we found that near to the cavity walls the fluctuations increase due to the screening of the electric field by the cavity. 
This approach was then connected to the study of Casimir and Van-der Waals forces, both of which are manifestations of forces induced by electromagnetic fluctuations. 

In the future it would be extremely interesting to extend this approach to include more complex heterostructures which feature multiple types material~\cite{Basov.20164a,Liu.2016} including metals~\cite{Jing.2021}, insulators~\cite{Caldwell.2019}, superconductors~\cite{Xi.2015,Yu.2019}, semiconductors~\cite{Amelio.2021}, ferroelectrics~\cite{Zheng.2020,Stern.2021,Yasuda.2021,Woods.2021,Moore.2021}, magnets~\cite{Rizzo.2022,Huang.2018,Lee.2021,Klein.2022,Klein.2022k3,Bedoya-Pinto.2021,Klein.2021}, and multiferroics~\cite{Song.2022}\textemdash all of which are phases of matter which can be characterized by their couplings to electromagnetic fields and which are now realizeable down to the two-dimensional limit~\cite{Novoselov.2016}.
In the future it would be very interesting to consider the mutual coupling of different stacked phases through their shared quantum electrodynamic environment. 
It is also interesting to extend these calculations into the ordered phase, where the order parameter as well as the fluctuations become important. 
Finally, probably the most important direction for future research are experimental realizations and confirmations of this theory.
To that end, it is likely necessary to use more accurately obtained parameters and electromagnetic solvers.
We therefore would envision interfacing this framework with {\it ab initio} calculations of microscopic parameters~\cite{Flick.2015,Flick.2019,Ruggenthaler.2014}, as well as finite-element Maxwell-equation solvers.
Conceptually, this is relatively straightforward but likely requires a large degree of technical work before it can be widely applied. 

\begin{acknowledgements}
We would like to acknowledge invaluable discussions with Yuto Ashida, Ankit Disa, Urs Staub, Prineha Narang, Ata\c{c} {\. I}mamo{\u g}lu, David Hsieh, Pavel Dolgirev, Aaron Lindenberg, N. Peter Armitage, John Sous, Maximilian Daschner, Andrea Cavalleri, J{\'e}r{\^o}me Faist, Dieter Jaksch, Misha Fogler, Ilya Esterlis, and John Philbin.
J.B.C. is an HQI Prize Postdoctoral Fellow and gratefully acknowledges support from the Harvard Quantum Initiative. 
ED acknowledges support from the ARO grant number W911NF-21-1-0184 and the SNSF project 200021\_212899.
\end{acknowledgements}
\bibliography{references}

\appendix

\section{Matsubara Formalism}
\label{app:matsubara}
In the Matsubara approach, we introduce the partition function via functional integral 
\begin{equation}
    Z = \int \mathcal{D}[Q,A,\chi] e^{-S[Q,A,\chi]}.
\end{equation}
The Matsubara action is written in the Weyl gauge with $\phi = 0$.
In principle this gauge is not unique and requires a further gauge fixing since the vector potential may still be shifted by a static gradient of the form $\mathbf{A} \to \mathbf{A} + \nabla \chi$ where $\chi$ is time independent.
However, we believe this remaining ambiguity does not cause problems in the treatment of the path integral.
Nevertheless, in the future a more rigorous and sophisticated method such as the Fadeev-Popov method should be employed. 

There is a subtlety about writing the electromagnetic field in the Matsubara formalism, which is that since the electric field $\mathbf{E}$ is in fact a canonical momentum (it is the time derivative of the coordinate $\mathbf{A}$), it carries an additional factor of $i$ when coupling to the polarization. 
Thus, the action is written as the integral of the Lagrangian 
\begin{multline}
   \mathcal{L} = \frac12\left[ \left( \frac{\partial \mathbf{A}}{\partial \tau} \right)^2 + \left( \nabla \times \mathbf{A} \right)^2 + \left( \frac{\partial \mathbf{Q}}{\partial \tau} \right)^2 + \Omega_0^2 \mathbf{Q}^2 \right] \\
   + i \eta \mathbf{Q}\cdot\frac{\partial \mathbf{A}}{\partial \tau} + i \frac12 \chi(x) \mathbf{Q}^2 + \frac{1}{4\lambda}\chi^2 
\end{multline}
over space and imaginary time $\tau \in [0,\beta]$.
Here we see the hybridization between the phonon and photon through the electric dipole coupling $\mathbf{Q}\cdot i\partial_\tau \mathbf{A} \to \mathbf{Q} \cdot\mathbf{E}$ upon returning to real time. 
Since the electromagnetic field experiences dispersion (due to the magnetic induction), it must also be subjected to boundary conditions which in this case are the same perfect-metal conditions we used previously. 

The last term is a Hubbard-Stratonovich term used to decouple the quartic phonon-phonon interaction in the Hartree channel.
Here the factor of $i$ reflects the interaction is repulsive. 
Integration over $\chi$ can be performed, and this returns the theory to its standard form in terms of $\mathbf{A}$ and $\mathbf{Q}$ with a nonlinearity $\sim \lambda (\mathbf{Q}^2 )^2$.
Qualitatively, $\chi$ represents the renormalization of the phonon frequency from bare value $\Omega_0$ to the physical value $\Omega_T$, which is renormalized by the phonon-phonon interactions. 

We now show that up to one-loop order this also yields the same result as the FDT calculation above. 
In particular, we expect the saddle-point in $\chi$ to correspond to the self-consistent Hartree approximation.
We can obtain this formally exactly by integrating out the phonon and photon, which now appear quadratically, to get
\begin{equation}
    \frac{\beta}{2\lambda} \chi(x) +\frac{\delta W[\chi]}{\delta \chi(x)} = 0
\end{equation}
with functional determinant
\begin{equation}
    W[\chi] = \frac12 \Tr \log \mathbb{K}[A]
\end{equation}
where the kernel can be identified as the Gaussian part of the Matsubara action.

This expression is simplified if we make an ansatz that the self-energy $\chi$ is time-independent in the saddle-point. 
Then, we can perform a shift to ``complete the square" for $\mathbf{Q}$, writing in the frequency domain 
\begin{equation}
    \mathbf{Q}(x) = \delta\mathbf{Q}(x) - \frac{\eta \omega_m}{\omega_m^2 + \Omega_0^2 + i\chi(\mathbf{r})}\mathbf{A}(x) .
\end{equation}
The first term characterizes the ``unscreened" fluctuations, as we argued in the first section, while the second term describes the screening due to the electromagnetic field. 
By writing $\chi(\mathbf{r})$ we have allowed for the possibility of an inhomogeneous shift in the phonon self-energy, as we expect near the boundary of the system. 

With this transformation the kernel decouples into the part due to $\delta \mathbf{Q}(x)$ and the part due to the dielectric energy. 
We also see that the variation with respect to $\chi$ can be framed as a variation with respect to the dressed phonon frequency, since they are related by 
\begin{equation}
    \Omega_T^2(\mathbf{r}) = \Omega_0^2 + i \chi(\mathbf{r}) \Rightarrow \frac{\delta }{\delta \chi(\mathbf{r})} = i\frac{\delta}{\delta(\Omega_T^2(\mathbf{r}))}.
\end{equation}
We find, after performing the transformation that the functional has two contributions;
\begin{equation}
    W_0[\chi] = \frac12 \sum_{\omega_m} \Tr \log \left[ \mathds{1}\delta^3(\mathbf{r}'-\mathbf{r}) \left( \omega_m^2 + \Omega_T^2(\mathbf{r}) \right) \right] 
\end{equation}
from the unscreened response (in the absence of phonon dispersion this is purely local and diverges with UV cutoff as $\Lambda^d$), and 
\begin{equation}
    W_{scr}[\chi] = \frac12 \sum_{\omega_m} \Tr \log \left[ \delta^3(\mathbf{r}'-\mathbf{r}) \left(\epsilon(i\omega_m,\mathbf{r})\omega_m^2 \mathds{1}- \nabla^2 + \nabla \nabla \cdot\right) \right]^{-1}
\end{equation}
from the dielectric screening. 
Now, all of the dependence on $\chi$ is captured through 
\begin{equation}
   \epsilon(i\omega_m,\mathbf{r}) = 1 + \frac{\eta^2}{\omega_m^2 + \Omega_T^2(\mathbf{r}) }.
\end{equation}
We therefore can easily evaluate the derivative in terms of the Matsubara Green's functions (note the minus sign is different than the usual definition here)
\begin{subequations}
\begin{align}
    & \mathscr{D}_0(\mathbf{r},\mathbf{r}'; \omega_m) = \left[ \mathds{1} \delta^3(\mathbf{r}'-\mathbf{r}) \left( \omega_m^2 + \Omega_T^2(\mathbf{r}) \right) \right]^{-1} \\
    & \mathscr{G}(\mathbf{r},\mathbf{r}'; \omega_m) = \left[ \delta^3(\mathbf{r}'-\mathbf{r}) \left(\epsilon(i\omega_m,\mathbf{r})\omega_m^2 \mathds{1} - \nabla^2 + \nabla \nabla \cdot\right) \right]^{-1} \\
\end{align}
\end{subequations}
as 
\begin{equation}
    \frac{\delta W[\chi]}{\delta \chi(\mathbf{r})} = i\frac12 \sum_{\omega_m} \tr \left[ \mathscr{D}_0(\mathbf{r},\mathbf{r}; \omega_m) + \omega_m^2 \frac{\delta \epsilon(i\omega_m)}{\delta\Omega_T^2} \mathscr{G}(\mathbf{r},\mathbf{r};\omega_m) \right].
\end{equation}
We therefore find an equation for $i\chi(\mathbf{r}) = \Omega_T^2(\mathbf{r}) -\Omega_0^2$ of 
\begin{multline}
        \frac{1}{\lambda}\left( \Omega_T^2(\mathbf{r}) - \Omega_0^2 \right) \\
        =  T \sum_{\omega_m} \tr \left[ \mathscr{D}_0(\mathbf{r},\mathbf{r}; \omega_m) + \omega_m^2 \frac{\delta \epsilon(i\omega_m,\mathbf{r})}{\delta\Omega_T^2} \mathscr{G}(\mathbf{r},\mathbf{r};\omega_m) \right]. 
\end{multline}
$\Omega_0^2$ is a counter term which is set by the renormalization condition that $\Omega_T^2(\mathbf{r})$ match the bulk value at a given temperature.

We now recover the result from the previous section if we evaluate the right-hand side to lowest order (i.e. not self-consistently) in $\lambda$, taking $\Omega_T^2(\mathbf{r}) = \Omega_{TO}^2$ to be the bulk TO mode frequency. 
Then the right-hand side is nothing but $\langle \mathbf{Q}(\mathbf{r},t)^2 \rangle $ evaluated as a Matsubara sum. 
This can now be evaluated efficiently, and in an unbiased manner, provided one can evaluate the Green's functions. 

\section{Variational Approach}
\label{app:variational}
As a final sanity check, we also provide a derivation based on a variational approach for the thermodynamic free energy. 
We again work in Matsubara formalism, but instead of using a Hubbard-Stratonovich transformation we employ the Feynman-Gibbs-Bogoliubov inequality.
To this end, we write the partition function as 
\begin{equation}
    Z = \int \mathcal{D}[\mathbf{Q},\mathbf{A}] e^{-S} = \int \mathcal{D}[\mathbf{Q},\mathbf{A}] e^{-S_{\rm eff}} e^{- (S - S_{\rm eff})}.
\end{equation}
The effective action is then chosen to constitute the variational ansatz.
We use 
\begin{multline}
    S_{\rm eff} = \int d^3 r \sum_{\omega_m} \left[\frac12 \mathbf{Q}_{-\omega}(\omega_m^2 + \Omega_{T}^2(\mathbf{r}) )\mathbf{Q}_{\omega} +\eta \omega_m \mathbf{Q}_{-\omega}\cdot\mathbf{A}_{\omega} \right]\\
    +S_{\rm Maxwell},
\end{multline}
which essentially replaces the bare TO frequency and interactions with a local, effective TO frequency.
Note this differs from the analysis of Ref.~\cite{Ashida.2020} so far only in the choice of a local variational TO frequency as opposed to a single global parameter. 
With respect to this ansatz, correlation functions may be calculated easily using the above frameworks, since the effective action used in the ansatz is quadratic and time independent. 

We then obtain an upper bound on the free energy as a functional of our ansatz via the Feynman-Bogoliubov-Gibbs inequality as
\begin{equation}
    W[\Omega_{T}^2(\mathbf{r})] \leq W_{\rm eff} + \langle S - S_{\rm eff} \rangle_{\rm eff}.
\end{equation}
The first term is simply the free energy of the noninteracting ansatz, while the second term becomes 
\begin{equation}
    \langle S - S_{\rm eff} \rangle_{\rm eff} = \int d^4x \left[ \frac{\lambda}{4} \langle (\mathbf{Q}^2(x))^2\rangle + \frac12(\Omega_{0}^2 - \Omega_T^2(\mathbf{r}) ) \langle \mathbf{Q}^2(x)\rangle \right].
\end{equation}
This can be evaluated using Wick's theorem. 
We now vary the parameter $\Omega_T^2(\mathbf{r})$ to find the best approximation to the free energy. 
The evaluation is aided by performing a canonical transformation, as done in Appendix~\ref{app:matsubara}. 
We shift $\mathbf{Q} = \delta \mathbf{Q} - \frac{\eta \omega_m}{\omega_m^2 + \Omega_T^2(\mathbf{r})} \mathbf{A}$. 
This then allows us to express the free energy derivative as 
\begin{equation}
    \frac{\delta W_{\rm eff}}{\delta \Omega_T^2(\mathbf{r})} = \frac{1}{2}\sum_{\omega_m} \tr \mathscr{\hat{D}}_0(\mathbf{r},\mathbf{r};\omega_m) + \frac{\delta \epsilon(i\omega_m,\mathbf{r})}{\delta \Omega_T^2(\mathbf{r})} \omega_m^2 \tr \mathscr{\hat{G}}(\mathbf{r},\mathbf{r};\omega_m) 
\end{equation}
where the two Green's functions are 
\begin{equation}
   \mathscr{\hat{D}}_0(\mathbf{r},\mathbf{r}';\omega_m) = \langle \delta \mathbf{Q}_{-\omega}(\mathbf{r}) \delta\mathbf{Q}_{\omega}(\mathbf{r}') \rangle_{\rm eff}
\end{equation}
and 
\begin{equation}
  \mathscr{\hat{G}}(\mathbf{r},\mathbf{r}';\omega_m) = \langle \mathbf{A}_{-\omega}(\mathbf{r})\mathbf{A}_{\omega}(\mathbf{r}')\rangle_{\rm eff}
\end{equation}
and the dielectric constant is found as 
\begin{equation}
    \epsilon(i\omega_m,\mathbf{r}) = 1 + \frac{\eta^2}{\omega_m^2 + \Omega_T^2(\mathbf{r})}.
\end{equation}
We note as well that by direct examination, we have 
\begin{equation}
 \frac{\delta W_{\rm eff}}{\delta \Omega_T^2(\mathbf{r})} = \int d\tau \frac12 \langle \mathbf{Q}^2(\mathbf{r})\rangle_{\rm eff} = \frac{1}{2}\sum_{\omega_m}\langle \mathbf{Q}_{-\omega}(\mathbf{r}) \cdot\mathbf{Q}_{\omega}(\mathbf{r})\rangle_{\rm eff},
\end{equation}
which implies that the local TO frequency is a variational parameter which is conjugate to the local phonon fluctuations.

We can now prove that this is equivalent to the previous two approaches. 
We first use Wick's theorem to derive the quartic term in terms of the quadratic propagator. 
In the isotropic, paraelectric phase, we have 
\begin{equation}
   \langle Q^a(x)Q^aQ^b(x)Q^b(x) \rangle =  \left( \langle \mathbf{Q}(x)^2 \rangle \right)^2 + \frac{2}{d}\left( \langle \mathbf{Q}(x)^2 \rangle \right)^2,
\end{equation}
where $d$ is the spatial dimension.
In the large $d$ limit this is simply the Hartree term, with the Fock term being subleading in that limit.

We can now write our variational functional, using the expression for the interaction derived above and the relation between $\langle \mathbf{Q}(x)^2 \rangle$ and the derivative of $W_{\rm eff} $ to get the exact result   
\begin{widetext}
\begin{equation}
    W = W_{\rm eff} + \int d^3 r' (\Omega_0^2 - \Omega_T^2(\mathbf{r})) \frac{\delta W_{\rm eff}}{\delta \Omega_T^2(\mathbf{r}')} + \frac{u}{\beta}(1 + \frac{2}{d}) \left( \frac{\delta W_{\rm eff}}{\delta \Omega_T^2(\mathbf{r}')}\right)^2  . 
\end{equation}
\end{widetext}
Now, we take a derivative with respect to the parameter $\Omega_T^2(\mathbf{r})$.
After applying the chain rule, one finds 
\[
\frac{\delta W}{\delta \Omega_T^2(\mathbf{r})} = F- F + \left[ \Omega_0^2 - \Omega_T^2(\mathbf{r}) \right]\frac{\delta F}{\delta \Omega_T^2} + 2\frac{u}{\beta}(1+ \frac{2}{d}) F \frac{\delta F}{\delta \Omega_T^2},
\]
where $F = \frac{\delta W_{\rm eff}}{\delta \Omega_T^2(\mathbf{r})}$ is the local phonon fluctuation density.
The first two terms cancel and the derivative of the phonon density is in general dependent on the TO frequency, so this leaves the variational equation simplified as 
\begin{equation}
     \left[ \Omega_0^2 - \Omega_T^2(\mathbf{r}) \right] + 2 u T (1 + \frac{2}{d}) \frac{\delta W_{\rm eff}}{\delta \Omega_T^2(\mathbf{r})} = 0 .
\end{equation}
We note that other than the factor of $2/d$ due to the Fock correction at finite $d$, this exactly matches our condition from the Hubbard-Stranovich method.

\section{High-Temperature Behavior}
\label{app:high-temperature}
Here we provide another argument for why the transverse electromagnetic modes decouple at high-temperature.
We begin with the quantum finite-temperature Lagrangian describing the phonons and their coupling to the electromagnetic field in a gauge independent form
\begin{widetext}
\begin{equation}
    \mathscr{L}  = \frac12 \left[ \left( \frac{\partial \mathbf{Q}}{\partial \tau}\right)^2 + \Omega_0^2 \mathbf{Q}^2 \right] + \frac12\left[ \left(\nabla \phi + \frac{\partial \mathbf{A}}{\partial \tau} \right)^2 + (\nabla \times \mathbf{A})^2  \right] + i\eta\mathbf{Q}\cdot \left(\frac{\partial \mathbf{A}}{\partial \tau} + \nabla \phi \right).
\end{equation}
We can make the argument more clear if we do partial integration on the coupling term, so express it directly in terms of the gauge fields
\begin{equation}
    \mathscr{L}  = \frac12 \left[ \left( \frac{\partial \mathbf{Q}}{\partial \tau}\right)^2 + \Omega_0^2 \mathbf{Q}^2 \right] + \frac12\left[ \left(\nabla \phi + \frac{\partial \mathbf{A}}{\partial \tau} \right)^2 + (\nabla \times \mathbf{A})^2  \right] - i\eta\left[ \phi \nabla \cdot\mathbf{Q} + \mathbf{A} \cdot \frac{\partial \mathbf{Q}}{\partial \tau} \right].
\end{equation}
\end{widetext}
We choose the Coulomb gauge, where $\nabla \cdot \mathbf{A} = 0$, such that $\mathbf{A}$ clearly couples to the {\bf transverse} currents induced by the phonon modes, as these are the true radiative degrees of freedom. 

We can then separate the longitudinal and transverse parts of the radiation coupling as 
\begin{equation}
    \mathscr{L}^{\parallel}_{\rm int}  = \frac12\left(\nabla \phi\right)^2 -i\eta \phi \nabla \cdot\mathbf{Q}^\parallel .
\end{equation}
This clearly endows the LO phonon mode with the long-range Coulomb interaction. 
It also clearly couples to the phonon coordinate $\mathbf{Q}$ and makes perfect sense in the classical limit, where $\mathbf{Q}$ and $\phi$ become ``time-independent" and the integral over Matsubara time becomes simply $\int_0^\beta d\tau \to 1/T$.

On the other hand, the transverse modes couple via 
\begin{equation}
    \mathscr{L}^{\perp}_{\rm int}  = \frac12\left(\frac{\partial \mathbf{A}}{\partial \tau} \right)^2 +\frac12 (\nabla\times\mathbf{A})^2-i\eta \mathbf{A}\cdot \frac{\partial \mathbf{Q}^\perp}{\partial \tau}.
\end{equation}
We note a number of important points.
The first is that the radiative electric field has both retardation (due to the first term), and intrinsic dynamics due to the inductive response of the second term.
Second, we see that the field couples to the transverse phonon {\bf current}.
In the classical limit high-temperature limit the retardation due to $(\partial_\tau \mathbf{A})^2$ goes away as the cost of a thermal tunneling event $\sim T$ becomes suppressed, and we are left with 
\begin{equation}
    \mathscr{L}^{\perp}_{\rm int}  = \frac12 (\nabla\times\mathbf{A})^2-i\mathbf{A}\cdot \mathbf{J}^\perp.
\end{equation}
In this case the current $\mathbf{J}^\perp = \eta \partial_\tau \mathbf{Q}$ is essentially the canonical momentum associated to the dielectric polarization, but at high-temperatures this becomes uncorrelated with $\mathbf{Q}$.
Therefore, the photon field indeed dresses the expectation value for the phonon current fluctuations $\langle (\mathbf{J}^\perp )^2 \rangle$, but this now is independent of the phonon fluctuations $\langle \mathbf{Q}^2 \rangle $.
In this way, we see that this is a quantum effect since the photons only couple to the phonon displacement through the commutation relations between the phonon displacement and current $[\mathbf{Q},\mathbf{J}]\neq 0$, and at high-temperatures this vanishes.

\section{Evaluation of Green's Function in Cavity}
\label{app:solution}

Here we elaborate on the details for the calculations of the Green's function in the slab geometry.
This largely follows Ref.~\cite{AGD}, which calculates nearly the same quantity we need but only evaluates at the boundary of the Fabry-Perot geometry; we want to evaluate it locally in the bulk. 
We must obtain the Matsubara Green's function for the vector potential in the Weyl gauge.
This satisfies the equation 
\begin{equation}
\left[ \omega_m^2 \epsilon(i\omega_m) + \begin{pmatrix}
- \partial_z^2 & 0 & iq \partial_z \\
0 & q^2 - \partial_z^2 & 0 \\
iq \partial_z & 0 & q^2 \end{pmatrix} \right] \mathbb{\hat{G}}(z,z') =  \delta(z-z') 
\end{equation}
where $q$ is the in-plane momentum, which we have taken to lie along $\mathbf{\hat{e}}_x$ (this can be done provided there is in-plane isotropy.

Clearly, the problem decoules in to the transverse electric (TE) modes, which solve 
\begin{equation}
\left[ \omega_m^2 \epsilon(i\omega_m) + q^2 - \partial_z^2 \right] G_{yy}(z,z') = \delta(z-z')
\end{equation}
(they have the electric field polarized along $\mathbf{\hat{e}}_y$ which is transverse to the momentum), and the transverse magnetic (TM) and longitudinal (L) modes which are coupled and solve 
\begin{equation}
\left[ \omega_m^2 \epsilon(i\omega_m) + \begin{pmatrix}
- \partial_z^2 &  iq \partial_z \\
iq \partial_z & q^2 \end{pmatrix} \right] \mathbb{\hat{G}}(z,z') =  \delta(z-z') .
\end{equation}
We also have boundary conditions on the $x,y$ components at $\pm L/2$. 

We begin with the TE modes.
These can be solved for analytically by the method of matching. 
We can write down the solution almost immediately (after some deep introspection)
\begin{equation}
G_{yy}(z,z') = \begin{cases}
A \sinh \kappa (L/2 -z) \sinh \kappa (z'+L/2) & z> z' \\ 
A \sinh \kappa (L/2 -z') \sinh \kappa (z+L/2) & z < z'. \\ 
\end{cases}
\end{equation}
We have introduced $\kappa = \sqrt{ \omega_m^2 \epsilon(i\omega_m) + q^2}$.
What now remains is to impose continuity of the derivative at $z = z'$. 
Integrating across the singularity and utilizing hyperbolic trig identities we obtain the relation 
\begin{equation}
A \kappa \sinh\kappa L = 1 . 
\end{equation}
We therefore obtain the TE mode Green's function in analytical form of 
\begin{multline}
G_{yy}(z,z') = \frac{1}{\kappa \sinh \kappa L }\\
\times \begin{cases}
 \sinh \kappa (L/2 -z) \sinh \kappa (z'+L/2) & z> z' \\ 
\sinh \kappa (L/2 -z') \sinh \kappa (z+L/2) & z < z'. \\ 
\end{cases}
\end{multline}

The case of the remaining two modes is harder, in particular since the equations are second order and coupled, allowing for the potential for a fourth order equation upon decoupling. 
Let us write out all the equations explicitly
\begin{widetext}
\begin{subequations}
\begin{align}
& (\omega_m^2 \epsilon(i\omega_m) - \partial_z^2 ) G_{xx}(z,z') + iq \partial_z G_{zx}(z,z') = \delta(z-z') \\
& \kappa^2 G_{zz}(z,z') + iq \partial_z G_{xz}(z,z') = \delta(z-z') \\
& (\omega_m^2 \epsilon(i\omega_m) - \partial_z^2 ) G_{xz}(z,z') + iq \partial_z G_{zz}(z,z') = 0 \\
& iq \partial_z G_{xx}(z,z') +\kappa^2 G_{zx}(z,z') = 0 .
\end{align}
\end{subequations}
\end{widetext}

We can locally eliminate $G_{zx}$ to get 
\begin{equation}
G_{zx}(z,z') = -\frac{1}{\kappa^2} iq \partial_z G_{xx}(z,z')
\end{equation}
giving 
\begin{equation}
(\omega_m^2 \epsilon(i\omega_m) - \partial_z^2 + \frac{q^2}{\kappa^2} \partial_z^2 )G_{xx}(z,z') =  \delta(z-z') 
\end{equation}
for the in-plane component.
This simplifies slightly to produce 
\begin{equation}
( \kappa^2 -  \partial_z^2 )G_{xx}(z,z') = \frac{\kappa^2 }{\omega_m^2\epsilon(i\omega_m)} \delta(z-z') 
\end{equation}
In fact, this can be solved just as trivially as the TE modes since we need only replace the constant prefactor $A$. 
We obtain 
\begin{multline}
G_{xx}(z,z') = \frac{\kappa}{ \omega_m^2\epsilon(i\omega_m) \sinh \kappa L }\\
\times \begin{cases}
 \sinh \kappa (L/2 -z) \sinh \kappa (z'+L/2) & z> z' \\ 
\sinh \kappa (L/2 -z') \sinh \kappa (z+L/2) & z < z'. \\ 
\end{cases}
\end{multline}
All in all we then obtain for the contributions $G_{xx}(z,z) + G_{yy}(z,z)$ together 
\begin{multline}
\tr \mathbb{G}_{\parallel}(z,z) \\
=\frac{\sinh\kappa(L/2 - z) \sinh\kappa(L/2 + z) }{\sinh\kappa L }\left[ \frac{1 }{\kappa } + \frac{\kappa}{\omega_m^2 \epsilon(i\omega_m)}\right].
\end{multline}
While this is not itself singular, we are reminded that we still must perform the integral over $q$ and this will in general incur a cutoff dependence. 

The last step is the most difficult; we must determine the normal component of $G$. 
The normal component is found following Ref.~\cite{AGD}, whereby we first obtain the off-diagonal component 
\begin{equation}
 G_{zx}(z,z') = -\frac{iq}{\kappa^2}\partial_z G_{xx}(z,z').
\end{equation}
It then seems that an assumption has been made in Ref.~\cite{AGD}, which seems we must make as well, which is that the Green's function is reciprocal in the sense that we can interchange $G_{zx}$ and $G_{xz}$, so as to close the equations. 
This seems valid provided time-reversal, or more importantly, reciprocity of the system is preserved. 

If we make this simplification, then we can obtain $G_{zz}$ from $G_{zx}$, ultimately giving  
\begin{equation}
\kappa^2 G_{zz}(z,z') =  (\omega_m^2 \epsilon(i\omega_m) - \partial_z^2)G_{xx}(z,z').
\end{equation}
We can in turn utilize the equation for $G_{xx}$ to remove the derivative, which is very singular acting on the Green's function. 
We then instead recover an inhomogeneous equation of 
\begin{equation}
\kappa^2 G_{zz}(z,z') =   - q^2 G_{xx}(z,z') + \frac{\kappa^2}{\omega_m^2 \epsilon(i\omega_m)} \delta(z-z'),  
\end{equation}
such that we have 
\begin{equation}
G_{zz}(z,z') =   -\frac{q^2}{\kappa^2} G_{xx}(z,z') + \frac{1}{\omega_m^2 \epsilon(i\omega_m)} \delta(z-z').
\end{equation}
In fact, the last term is the origin of the most severe singularity, since the Green's function itself is singular at $z= z'$, not merely non-differentiable. 
However, we are saved by the fact that this term is essentially constant and independent of the geometry. 
We therefore obtain the complete expression for the local vector-potential fluctuations as 
\begin{widetext}
\begin{multline}
\tr \mathbb{G}(z,z) =  \frac{\sinh\kappa(L/2 - z) \sinh\kappa(L/2 + z) }{\sinh\kappa L }\\
\times\left[ \frac{1 }{\kappa } + \frac{\kappa}{\omega_m^2 \epsilon(i\omega_m)} - \frac{q^2}{\kappa\omega_m^2 \epsilon(i\omega_m)} \right] + \frac{1}{\omega_m^2 \epsilon(i\omega_m) }\delta(0).
\end{multline}
\end{widetext}
The quantity in brackets can be simplified as 
\begin{equation}
\left[ \frac{1}{\kappa } + \frac{\kappa}{\omega_m^2 \epsilon(i\omega_m)} - \frac{q^2}{\kappa\omega_m^2 \epsilon(i\omega_m)} \right]  = \frac{2}{\kappa}.
\end{equation}
We therefore obtain the result for the electric-field fluctuations, which are weighted by an additional $\omega_m^2$ factor. 
This gives 
\begin{multline}
\omega_m^2 \tr \mathbb{G}(z,z) =\\
\frac{ \omega_m^2}{2\kappa}\frac{\sinh\kappa(L/2 - z) \sinh\kappa(L/2 + z) }{\sinh\kappa L } + \frac{1}{ \epsilon(i\omega_m) }\delta(0).
\end{multline}
We now must subtract the part which is independent of geometry or system size.
In particular, what matters for the Casimir formula is 
\begin{equation}
\Delta \langle E E \rangle = \omega_m^2 \left[ \tr \mathbb{G}(z,z) - \tr \mathbb{G}(0,0)\bigg|_{L\to \infty} \right].
\end{equation}
We find a simple result of 
\begin{equation}
\Delta \langle E E \rangle =  \frac{\omega_m^2}{2\kappa}\left( \frac{\sinh\kappa (L/2 - z) \sinh \kappa (L/2 +z)}{\sinh \kappa L} - \frac12  \right).
\end{equation}
Therefore, the result ends up being relatively simple (we still do need to integrate over $q$ and sum over $\omega_m$, to be clear).

We can already see however that this is going to be positive comparing the bulk and surface. 
In particular, we have 
\begin{equation}
\Delta \langle E E \rangle(L/2) - \Delta \langle EE\rangle(0) =  -\frac{\omega_m^2}{2\kappa} \sinh^2( \kappa L/2) /\sinh\kappa L .
\end{equation}
This is therefore manifestly negative; now if we recall that the overall contribution goes as $-\langle E^2\rangle$ due to the change in dielectric constant being inverse to the phonon frequency, we find that this will lead to a mode hardening at the boundary. 

In total, we find that the final result including the Matsubara sum and momentum integral is 
\begin{widetext}
\begin{equation}
\Delta \Omega_T^2(z) = \lambda T \sum_{\omega_m} \int^\Lambda \frac{d^2q}{(2\pi)^2} \frac{\partial \epsilon(i\omega_m)}{\partial \Omega_T^2} \frac{\omega_m^2}{2\kappa}\left( \frac{\sinh\kappa (L/2 - z) \sinh \kappa (L/2 +z)}{\sinh \kappa L} - \frac12  \right).
\end{equation}
\end{widetext}
This is to be numerically evaluated.
We do need to be careful since the integrand is singular in small $q$, $\omega$. 

We therefore numerically compute 
\begin{widetext}
\begin{multline}
\Delta \Omega_T^2(z) = \frac{\lambda }{2\pi }T \sum_{m = 1}^{\frac{\omega_c}{2\pi T}} \left[ - \frac{\omega_m^2 \eta^2}{(\omega_m^2 + \Omega_T^2)^2} \right]  \int_{\sqrt{\omega_m^2 \epsilon(i\omega_m)} }^{\sqrt{\omega_m^2 \epsilon(i\omega_m) + \Lambda^2}} d\kappa \frac{1}{2\kappa }\left( \frac{\sinh\kappa (L/2 - z) \sinh \kappa(L/2 +z)}{\sinh \kappa L} - \frac12  \right) \\ 
 + \frac{\lambda}{4 \pi T} \left[ -\frac{\eta^2}{\Omega_T^4} \right] \lim_{\omega \to 0} \omega^2\int_{\omega \sqrt{\epsilon(0)} }^{\Lambda} d\kappa \frac{1}{2\kappa }\left( \frac{\sinh\kappa (L/2 - z) \sinh \kappa(L/2 +z)}{\sinh \kappa L} - \frac12  \right).
\end{multline}
\end{widetext}
The first terms are the quantum corrections, which are evaluated by summing over the $m= \pm 1,\pm 2,...$ Matsubara frequencies, and this can be evaluated relatively easily numerically.
The last term is the classical contribution which is tricky to evaluate due to the singular limit as $\omega \to 0$. 
the last term does indeed vanish in this limit.
This is seen by first manipulating the integrand via $x = \kappa L $ and utilizing trig identities to get 
\begin{widetext}
\begin{equation}
\lim_{\omega \to 0} \omega^2\int_{\omega \sqrt{\epsilon(0)} }^{\Lambda} d\kappa \frac{1}{2\kappa }\left( \frac{\sinh\kappa (L/2 - z) \sinh \kappa(L/2 +z)}{\sinh \kappa L} - \frac12  \right) =\lim_{\omega \to 0} \frac{\omega^2 }{4} \int_{L \omega \sqrt{\epsilon(0)}}^{L\Lambda} \frac{dx}{x} \left[ -1 + \frac{\sinh x(\frac12 - s)\sinh x (\frac12 + s)}{\sinh x} \right].
\end{equation}
\end{widetext}
We can now apply L'H{\^o}spital's rule to this to determine the result. 
Applying this once we find 
\begin{widetext}
\begin{equation}
\lim_{\omega \to 0} \frac{\omega^2}{4} \int_{L \omega \sqrt{\epsilon(0)}}^{L\Lambda} \frac{dx}{x} \left[ -1 + \frac{\sinh x(\frac12 - s)\sinh x (\frac12 + s)}{\sinh x} \right] = \lim_{\omega \to 0} \frac{1}{2 /\omega^3} \frac14 \frac{1}{x} \left[ -1 + \frac{\sinh x(\frac12 - s)\sinh x (\frac12 + s)}{\sinh x} \right]\bigg|_{x= L\sqrt{\epsilon(0)}\omega}.
\end{equation}
\end{widetext}
The first term in the brackets is non-singular so we find 
\begin{widetext}
\begin{equation}
\lim_{\omega \to 0} \frac{\omega^2}{4} \int_{L \omega \sqrt{\epsilon(0)}}^{L\Lambda} \frac{dx}{x} \left[ -1 + \frac{\sinh x(\frac12 - s)\sinh x (\frac12 + s)}{\sinh x} \right] = \lim_{\omega \to 0} \frac{\omega^2}{8}\frac{1}{ L \sqrt{\epsilon(0)}} \left[ \frac{\sinh x(\frac12 - s)\sinh x (\frac12 + s)}{\sinh x} \right]\bigg|_{x= L\sqrt{\epsilon(0)}\omega}.
\end{equation}
\end{widetext}
To proceed further we write 
\[
\sinh x(\frac12 - s)\sinh x (\frac12 + s) = \frac12 ( \cosh x - \cosh 2 x s ) .
\]
Once again, the first term is not singular enough to possibly cancel the $\omega^2$ numerator. so we find the only possible way out is from evaluating 
\begin{widetext}
\begin{equation}
\lim_{\omega \to 0} \frac{\omega^2}{4} \int_{L \omega \sqrt{\epsilon(0)}}^{L\Lambda} \frac{dx}{x} \left[ -1 + \frac{\sinh x(\frac12 - s)\sinh x (\frac12 + s)}{\sinh x} \right] = \lim_{\omega \to 0} \frac{\omega^2}{16}\frac{1}{ L \sqrt{\epsilon(0)}} \frac{\cosh 2 \omega \sqrt{\epsilon(0)} z }{\sinh L\sqrt{\epsilon(0)} \omega }.
\end{equation}
\end{widetext}
Since $z$ is ultimately bounded by $L/2$ this also cannot scale with $z$ fast enough to possibly yield a singular limit, so we see that applying L'H\^ospitals rule once more we will find this limit vanishes. 
Thus, we can disregard the classical component altogether, since it does not contribute to this quantity. 

This establishes that the electrostatic fluctuations do not end up contributing, and we find that the result is simply due to the quantum fluctuations {\it vis a vis} 
\begin{widetext}
\begin{equation}
\Delta \Omega_T^2(z) = \frac{\lambda }{2\pi }T \sum_{m = 1}^{\frac{\omega_c}{2\pi T}} \left[ - \frac{\omega_m^2 \eta^2}{(\omega_m^2 + \Omega_T^2)^2} \right]  \int_{\sqrt{\omega_m^2 \epsilon(i\omega_m)} }^{\sqrt{\omega_m^2 \epsilon(i\omega_m) + \Lambda^2}} d\kappa \frac{1}{2\kappa }\left( \frac{\sinh\kappa (L/2 - z) \sinh \kappa(L/2 +z)}{\sinh \kappa L} - \frac12  \right) .
\end{equation}
\end{widetext}

\end{document}